\documentclass[12pt,a4paper]{article}
\usepackage[cp1251]{inputenc}
\usepackage[T2A]{fontenc}
\usepackage{amsmath} 
\usepackage{amsfonts}
\usepackage{graphicx}
\usepackage{amssymb}
\usepackage{xy}
\usepackage{multirow}

\newcounter{abcd}

\begin{document}
\title{Bell-type inequalities and upper bounds of multi-qudit states}
\author{Loran V. Akopyan, Vladimir I. Man'ko \\
\texttt{loran.akopyan@phystech.edu, manko@sci.lebedev.ru}}
\date{July 1, 2009}
\maketitle

\abstract{Multi-qudit systems are studied in tomographic probability representations of quantum qudit states. Results of calculations for Bell-type numbers within the framework of classical probability theory and in quantum tomography are compared. Violation of Bell-type inequalities are shown explicitly using the method of averaging in tomographic picture of quantum states.}

\section{Introduction}

The Bell-CHSH inequalities \cite{On the Einstein-Podolsky-Rosen Paradox}, \cite{Proposed experiment to test local hidden-variable theories} and their violation are considered as specific properties of non-local quantum correlations. On the other hand, recently \cite{Classical statistical distributions can violate Bell-type inequalities}, \cite{student-yad-fiz} it became clear that the Bell inequalities can be associated with properties of some probability distributions in classical probability theory \cite{feller}. From this point of view violation of Bell and Bell-type inequalities \cite{Bell inequalities for arbitrarily high dimensional systems}, \cite{New Bell inequalities for the singlet state: Going beyond the Grothendieck bound}, \cite{A Relevant Two Qubit Bell Inequality Inequivalent to the CHSH Inequality}, \cite{Clauser-Horne inequality for qutrits} can also be studied in the framework of standard probability theory. The aim of this work is to extend the results of our previous work \cite{Bell-type inequalities in classical probability theory} and establish Bell-type inequalities for random systems with specific joint-probability distributions. We show that there exist families of joint-probability distributions depending on extra parameters which violate Bell-type inequalities. We apply tomographic probability representation \cite{mancini}, \cite{An introduction to the tomographic picture of quantum mechanics} of quantum qudit states and show that the joint-probability distributions (tomograms) describing the multiqudit states  can violate Bell-type inequalities. The systems of two qubits, three qubits, two qutrits and qubit-qutrits are considered explicitly. The paper is organized as follows. In Section 2 we present the tomographic probability representation approach to studying multipartite systems. We discuss the origin of Bell-type inequalities and their relation to quantum non-local correlations. The violation of Bell-type inequalities is linked to the general problem of separability and entanglement of quantum states. We illustrate the theory on the well known example of two-qubit system and point out several cases of separable and entangled states based on the analysis of Bell-type inequality for the system and its upper bounds. In Section 3 we study the qubit-qutrit system in both classical and quantum tomographic probability representations of multipartite states. We establish both classical and quantum upper bounds for Bell-type inequalities in qubit-qutrit system. Several examples of parameters violating classical Bell-type inequalities are presented and their domains are calculated. Finally, in Sections 4 and 5 two-qutrit and three-qubit states are studied in terms of establishing the upper bounds of possible Bell-type inequalities analogous to Tsirelson bound in two-qubit case.

A set of non-negative real numbers normalized to unity  can be considered as some probability distribution $P(n)$ with a finite number of $N$ outcomes. One can then rearrange this set to receive yet another probability distribution $P(m)$. One can say that $P(m)$ is induced by $P(n)$. Moreover, some of these induced probability distributions can be interpreted as marginal distributions for the initial $P(n)$ which in its own turn can be regarded as a new joint-probability distribution $w_{ij}$. But then having a joint probability distribution one can associate it with the probability distribution of observables pertaining to measurements in corresponding subsystems and construct various functionals in the form of average values of these observables. However, it is clear that the fact of existence of such marginal and joint-probability distributions strongly depends on the properties of the initial set or table of probabilities. Thus, if they indeed exist, the functionals constructed as described above must by necessity provide some inequalities similar to Bell-CHSH or analogous constraints on joint-probability induced functions. It is in this context that one of the aims of our work is to show that and how Bell-type inequalities are associated with probability distributions or stochastic matrices deduced from the latter. The violation of Bell-type inequalities by Quantum mechanics is due to the specific properties of quantum tomograms inducing joint-probability distributions with special properties.

Our work follows conceptually the idea of possible generalization of Bell-type inequalities over quantum systems with arbitrary number of qudits similar to \cite{Extending Bell inequalities to more parties}. However, the method of tomographic description of quantum states \cite{Malkin and Manko book on coherent states} allows one to constitute and examine higher dimensional systems based only on the properties of some specially constructed probability vectors and stochastic matrices. The variety of possible Bell-type inequalities in a system made up of $s$ subsystems of $d_1,d_2,\ldots d_s$ dimensions accordingly, is conditioned by the choice of some special sign matrix denoted by our convention as $\hat{I}_{d_1\otimes d_2\otimes\ldots\otimes d_s}$. We develop a phenomenological method for constructing this matrix by considering the order of signs entering the expressions of both classical and quantum-wise averaged observables. In principle, the approach could allow to unite many concurring Bell-type inequalities within the framework of a standard probability theory. Through our work on the example of two-qutrit and three-qubit systems families of Bell-type inequalities are discovered which are not violated by quantum tomograms of entangled states. This results echo back to the well-known ``openness'' of the problem of establishing the widest constraints on correlations by local realistic theories \cite{Violations of local realism with quNits up to N=16}, \cite{Bell's theorem for general N-qubit states}. It then turns out that Bell-type inequalities and their upper bounds $B_{max}$ of multi-qudit states can be written in a general form by introducing a single correlation matrix $\hat{C}$ generalizing the special matrix $\hat{I}_{d_1\otimes d_2\otimes\ldots\otimes d_s}$. This allows to view $B_{max}$ as some general functions defined on the elements $c_{m_1,m_2,\ldots,m_s}$ of the matrix $\hat{C}$. This result is quite valuable to us as one can then construct multi-moment correlation functions in search of stronger Bell-type inequality violations.  Throughout our work we see that if the  Bell-type inequalities of an entangled multiqudit state are violated they tend to violate stronger than for two-qubit system \cite{Violations of local realism by two entangled quNits are stronger than for two qubits}, although the factual search of such states is in practice much more complicated, too. A general survey and discussion of Bell-type inequalities was presented in \cite{Bell inequalities many questions a few answers}.

\section{Observables and Correlations}
Let's examine the general approach of calculating averages and correlations in multipartite systems within the classical tomographic representation of states. For simplicity consider composite system of two qubits or two-level subsystems. In the past few years several new Bell-type inequalities were discovered \cite{A Relevant Two Qubit Bell Inequality Inequivalent to the CHSH Inequality} to pinpoint non-local variable theories violated by quantum mechanics. Our method has the additional advantage of tightening or loosing the Bell-type inequalities by the appropriate selection of a special matrix $\hat{C}$ which is inspired by recent findings of tight Bell-type inequalities \cite{Extending Bell inequalities to more parties}. Consider qubit as a quantum-mechanical spin $1/2$ system with only two internal degrees of freedom. A quantum-mechanical observable $C$ as a function of local observables which in this case are the spin projections $m$ can be represented by a $2\times2$ matrix in the standard spin $1/2$ basis $|m\rangle,m=\pm1/2$. Its quantum-mechanical average can be written in terms of classical probability theory:

\begin{displaymath}
\langle \hat{C}\rangle=\sum_{m,m'}c_{mm'}w_{m,m'}=\sum_{m,m'}\langle m'|\hat{C}^{\dag}|m\rangle\langle m|\hat{M}^{(\frac12)}_{w}|m'\rangle=Tr\left(\hat{C}^{\dag}\hat{M}^{(\frac12)}_{w}\right)=\langle \hat{C}|\hat{M}^{(\frac12)}_{w}\rangle
\eqno{\addtocounter{abcd}{1}(\arabic{abcd})}
\end{displaymath}

Here we have introduced the $2\times2$ stochastic matrix $\hat{M}^{(\frac12)}_{w}$ built up from tomographic probabilities $w(+\frac12, \hat{U})$ and $w(-\frac12, \hat{U})$ of qubit "up" ($m=+\frac12$) and "down" ($m=-\frac12$) states, $\hat{U}$ is the unitary rotation matrix entering the definition of the quantum tomogram \cite{Qubit-portraits of qudit states and quantum correlations}, \cite{Qubit portrait of qudit states and Bell inequalities}. From the other hand, given the density matrix $\hat{\rho}$ of the quantum state one can write for the average of $C$ the well known quantum-statistical formula:
\begin{displaymath}
\langle \hat{C}\rangle=Tr(\hat{\rho}\hat{C})=\langle\hat{\rho}|\hat{C}\rangle
\eqno{\addtocounter{abcd}{1}(\arabic{abcd})}
\end{displaymath}

A substantial difference between the two ways of determining the \emph{same} average value $\langle \hat{C}\rangle$ is that in the case of $(2)$ density matrix contains probabilities entering the final expression only via its diagonal elements whereas the matrix $\hat{M}_w$ is wholly built of probabilities. It is true that in case of a solitary qubit the difference is purely mathematical since there are only two probabilities $w(+\frac12)$ and $w(-\frac12)$ entering the final expression for the average value. However, in case of complex systems built up of several qubits the first method $(1)$ (i.e. the method based on determining the average of observables via stochastic matrices and probability vectors associated with the state $\hat{\rho}$) has a clear advantage over the method based on density matrix description of states as it allows the experimentator to view all the quantities involved as physical observables i.e. the matrices $\hat{C}$, quantum tomograms and $\hat{M}_w$. Moreover, if one wishes to measure average values of observables in a rotated coordinate systems in $(2)$ one needs to conduct $SU(2)$ transformation for both $\hat{\rho}$ and $\hat{C}$ i.e. write

\begin{displaymath}
\hat{\rho}\rightarrow \hat{U}^{\dag}\hat{\rho}\hat{U},\quad \hat{C}\rightarrow \hat{T}(\hat{U})\hat{C}\hat{T}(\hat{U}),
\end{displaymath}

where $\hat{T}(\hat{U})$ is a unitary transform acting on the original $\hat{C}$ matrix.

In quantum tomographic representation $(1)$ one only needs to indicate different directions in space entering the tomograms as column-vectors in matrix $\hat{M}_w$ (see below the matrix $(6)$). This allows to construct Bell-type inequalities and observe their violation in principle at an arbitrary number of directions. These are directions along which measurements can be carried out. Thus, to discover new Bell-type inequalities and upper bounds of multiqudit states one simply needs to set the correlation matrix $\hat{C}$ and construct the stochastic matrix $\hat{M}_w$. Let's illustrate the above said on the example of a system made up of only two qubits.

In two-qubit system one registers four possibilities for the composite state $w_{++}$, $w_{+-}$, $w_{-+}$, $w_{--}$ which can be arranged into a stochastic matrix:
\begin{displaymath}
\hat{M}^{(\frac12\frac12)}_{w}=
\left|\left| \begin {array}{cccc} w_{{++}}&w_{{++}}&w_{{++}}&w_{{++}}\\
\noalign{\medskip}w_{{+-}}&w_{{+-}}&w_{{+-}}&w_{{+-}}\\
\noalign{\medskip}w_{{-+}}&w_{{-+}}&w_{{-+}}&w_{{-+}}\\
\noalign{\medskip}w_{{--}}&w_{{--}}&w_{{--}}&w_{{--}}
\end {array}
\right|\right|
\eqno{\addtocounter{abcd}{1}(\arabic{abcd})}
\end{displaymath}
The correlations between dichotomic variables $m_1$ and $m_2$ inside the composite system are given then by formulas
\begin{displaymath}
\langle m_1m_2\rangle=\sum_{m_1,m_2}m_1m_2\cdot w_{m_1m_2}=Tr\left(\hat{\sigma}_{2\otimes2}^T\hat{M}_w\right)=\langle\sigma_{2\otimes2}|\hat{M}_w\rangle
\eqno{\addtocounter{abcd}{1}(\arabic{abcd})}
\end{displaymath}
where we have introduced the scalar product of operators in the standard spin $1/2$ basis $|m_1\rangle\otimes|m_2\rangle$. The matrix $\hat{\sigma}_{2\otimes2}$ consists of unity elements taken with the same signs with which each of the probabilities $w_{ij}$ enter the expression for correlations. It reads:
\begin{displaymath}
\hat{\sigma}_{2\otimes2}=\left|\left| \begin {array}{rrrr} 1&1&1&1\\\noalign{\medskip}-1&-1&-1&-1
\\\noalign{\medskip}-1&-1&-1&-1\\\noalign{\medskip}1&1&1&1\end {array}
\right|\right|
\eqno{\addtocounter{abcd}{1}(\arabic{abcd})}
\end{displaymath}
Since $w_{++}+w_{+-}+w_{-+}+w_{--}=1$, for the considered stochastic matrix $(3)$ one will always get the inequality $|\langle m_1m_2\rangle|\leq1$ which is a simple Cauchy-Schwartz inequality that the correlation cannot exceed $1$ in absolute value. In $(3)$ the four columns have the interpretation of discrete probability distributions. In order to predict non-local correlations, we must equip the theory with physically meaningful parameters which can be measured experimentally. Most commonly these are the Euler angles setting directions for two qubits to be oriented in space. Thus, varying the probability column-vectors entering $(3)$ along some four fixed directions $\mathbf{x},\mathbf{y},\mathbf{z},\mathbf{t}$ and assigning them different letters we arrive at a ``realistic'' stochastic matrix:
\begin{displaymath}
\hat{M}_{\mathbf{x},\mathbf{y},\mathbf{z},\mathbf{t}}=
\left|\left| \begin {array}{cccc} x_{{1}}&y_{{1}}&z_{{1}}&t_{{1}}
\\\noalign{\medskip}x_{{2}}&y_{{2}}&z_{{2}}&t_{{2}}
\\\noalign{\medskip}x_{{3}}&y_{{3}}&z_{{3}}&t_{{3}}
\\\noalign{\medskip}1-x_{{1}}-x_{{2}}-x_{{3}}&1-y_{{1}}-y_{{2}}-y_{{3}
}&1-z_{{1}}-z_{{2}}-z_{{3}}&1-t_{{1}}-t_{{2}}-t_{{3}}\end {array}
\right|\right|
\eqno{\addtocounter{abcd}{1}(\arabic{abcd})}
\end{displaymath}
Each of variables $x_i$, $y_j$, $z_k$, $t_l$, $i$, $j$, $k$, $l$ $=$ $1$, $2$, $3$ in $(6)$ can assume values between $0$ and $1$. The columns of this stochastic matrix represent four possible probability distributions of the two-qubit system. These probability column-vectors correspond to two distinct space orientation of each qubit in the composite system. In case of quantum probability distributions the variables $x_i$, $y_j$, $z_k$, $t_l$ become functions of $16$ Euler angles and thus render into continuous probability distributions\footnote{In reality there are six Euler angles attached to each tomographic probability distribution entering the resulting stochastic matrix as its columns. However, the azimuthal angles $\psi_n,n=1,2,3,4$ don't enter the final expressions. Thus the total number of angular parameters necessary to determine stochastic matrix associated with a given composite quantum state $\hat{\rho}$ reduces from $24$ to $16$.}.

The results $(3)-(6)$ allow generalization of our method in calculating correlations inside multipartite systems. The quantum-mechanical average of an observable $\hat{C}$ describing an expected correlation between local observables $m_1$ and $m_2$ (with matrix elements $c_{m_1m_2}=\langle m_1|\hat{C}|m_2\rangle$) is given by the formula:
\begin{gather*}
\langle \hat{C}\rangle=\sum_{m_1,m_2,m'_1,m'_2}c_{m_1m_2,m'_1m'_2}w_{m_1m_2,m'_1m'_2}=\sum_{m'_1,m'_2}\langle m'_1m'_2|\hat{C}^{\dag}\left(\sum_{m_1,m_2}|m_1m_2\rangle\langle m_1m_2|\right)\hat{M}^{(\frac12\frac12)}_{w}|m'_1m'_2\rangle=\\=Tr\left(\hat{C}^{\dag}\hat{M}^{(\frac12\frac12)}_{w}\right)=\langle \hat{C}|\hat{M}^{(\frac12\frac12)}_{w}\rangle
\end{gather*}
where the measurements of each of local observables $m,m'$ can yield up to $(2d+1)\times(2d'+1)$ outcomes, $d=m_{max},d'=m'_{max}$.

From these definition of correlations eligible both in the classical local (hidden) variable theories and the quantum mechanics it follows that an arbitrary \emph{linear} ``correlation form'' is the familiar average value:
\begin{displaymath}
B\left(\mathbf{x},\mathbf{y},\mathbf{z},\mathbf{t}\right)=Tr\hat{C}\hat{M}_{\mathbf{x},\mathbf{y},\mathbf{z},\mathbf{t}}
\eqno{\addtocounter{abcd}{1}(\arabic{abcd})}
\end{displaymath}
where the matrix $\hat{M}_{\mathbf{x},\mathbf{y},\mathbf{z},\mathbf{t}}$ is the stochastic matrix $(6)$ associated with the two-qubit system. The word "linear" refers to our way of calculation in standard probability theory as expressed in $(1)$. The form of $(6)$ is the most general expression for a stochastic matrix associated with two-qubit states both in classical and quantum cases. The building blocks for the theory are the matrices $\hat{\sigma}_{d_1\otimes d_2\otimes\ldots\otimes d_s}$ and $\hat{I}_{d_1\otimes d_2\otimes\ldots\otimes d_s}$ as concrete realizations of a general correlation matrix $\hat{C}_{d_1\otimes d_2\otimes\ldots\otimes d_s}$ violating Bell-type inequalities. The mechanism of constructing $\hat{\sigma}_{d_1\otimes d_2\otimes\ldots\otimes d_s}$ was described in \cite{Bell-type inequalities in classical probability theory}. The matrix $\hat{I}_{d_1\otimes d_2\otimes\ldots\otimes d_s}$ is a simple induction of the latter by transposition and sign inversion operations and was also defined there. The multivariate function $B\left(\mathbf{x},\mathbf{y},\mathbf{z},\mathbf{t}\right)$ reminds one the function that results in any of various Bell-CHSH inequalities and may be in that respect further named Bell-CHSH type\footnote{For simplicity we will be sometimes referring to the form $(7)$ as simply Bell-CHSH form.} form. It was established in \cite{Bell-type inequalities in classical probability theory}, \cite{An introduction to the tomographic picture of quantum mechanics} that in classical local variable theories the Bell-CHSH form is a harmonious function of its arguments. This formally results in Bell-type inequalities in classical probability theories. It is then natural to call the maxima of the function $B\left(\mathbf{x},\mathbf{y},\mathbf{z},\mathbf{t}\right)$ Bell-type numbers. In this paper we call the greatest in value among all-possible Bell-type numbers the upper-bound of the multi-qudit state of study. In composite systems the Bell-type numbers (and upper-bounds) are a unique but not complete measure of the strength of both local classical and non-local quantum correlations.

If the composite state is separable it contains no non-local correlations and the Bell-CHSH form is limited to some maximal value. In some cases the Bell-type numbers may even frequently vanish. Indeed, if we choose the matrix $\hat{C}$ equal to $\hat{\sigma}$ constructed from all-possible combinations of dichotomic variables $m_1$ and $m_2$ we notice that:
\begin{displaymath}
B\left(\mathbf{x},\mathbf{y},\mathbf{z},\mathbf{t}\right)=Tr\hat{\sigma}_{2\otimes2}\hat{M}_{\mathbf{x},\mathbf{y},\mathbf{z},\mathbf{t}}=\sum_ix_i-\sum_jy_j-\sum_kz_k+\sum_lt_l=0
\end{displaymath}
However, if the system is entangled the joint-probabilities entering the stochastic matrix $\hat{M}_{\mathbf{x},\mathbf{y},\mathbf{z},\mathbf{t}}$ don't factorize and the non-local quantum correlations may overthrow the classical upper bounds or maximal correlation strengths.

It is interesting to note, that by the special choice of matrix $\hat{C}$ the Bell-CHSH form reduces its dependence only to four parameters. Indeed, the direct calculation shows
\begin{displaymath}
B\left(\mathbf{x},\mathbf{y},\mathbf{z},\mathbf{t}\right)=Tr\hat{P}\hat{M}_{\mathbf{x},\mathbf{y},\mathbf{z},\mathbf{t}}=2\left(x_1+y_1+z_1-t_1-1\right)
\end{displaymath}
where the matrix $\hat{P}$ differs from an earlier matrix $\hat{I}_{2\otimes2}$ from \cite{Bell-type inequalities in classical probability theory} with the change of sign in the last column. Thus in the case of matrix $\hat{P}$ the composite state is described by the same number of parameters as in the case of separable two-qubit state. Had we taken the matrix $\hat{\sigma}_{2\otimes2}$ in form of $(5)$ we would have received all Bell-type numbers null just as the matrix associated with the joint-probability distribution would have shown. The Bell-type inequalities and upper bounds are received by examining the properties of the B-form in $(7)$. The minimums and maximums of the Bell-CHSH form are achieved only at the $2^4=16$ boundaries of probability distributions  $x_i$, $y_j$, $z_k$, $t_l$ in full agreement with an earlier similar approach described in \cite{An introduction to the tomographic picture of quantum mechanics}. All sixteen Bell-CHSH type inequalities are restrained in the classical domain of $|B|\leq2$ i.e. the upper bound in case of two qubits is $2$. In Section $7.1$ of Appendix the results of analytic and numeric computations for two qubit system are presented and here we would like to stop on detailed description of the analysis of the quantum case of entangled system of two qubits.

The tomogram of state described by a $2\times2$ density matrix $\hat{\rho}$ is presented by the probability vector
\begin{displaymath}
\mathbf{w}\left(\hat{\rho}, \hat{U}\right)=Diag\left(\hat{U}^{\dag}\hat{\rho}\hat{U}\right)=\left|\left|
\begin {array}{c}
|u_{11}|^2\rho_{11}+u_{11}\overline{u_{21}}\rho_{21}+u_{21}\overline{u_{11}}\rho_{12}+|u_{21}|^2\rho_{22}\\\\
|u_{12}|^2\rho_{11}+u_{12}\overline{u_{22}}\rho_{21}+u_{22}\overline{u_{12}}\rho_{12}+|u_{22}|^2\rho_{22}
\end {array}
\right|\right|
\end{displaymath}
Entering rotation operator matrix from $SU(2)$ in the general form, we arrive at
\begin{displaymath}
\mathbf{w}\left(\hat{\rho}, \theta, \varphi\right)=\left|\left|
\begin {array}{c}
\rho_{11}\cos^2{\dfrac{\theta}2}-\sin{\theta}Re\left(\rho_{12}e^{-i\varphi}\right)+\rho_{22}\sin^2\dfrac{\theta}2\\\\
\rho_{11}\cos^2{\dfrac{\theta}2}+\sin{\theta}Re\left(\rho_{12}e^{-i\varphi}\right)+\rho_{22}\cos^2{\dfrac{\theta}2}
\end {array}
\right|\right|
\end{displaymath}
where $\varphi$ and $\theta$ are the Euler angles entering the expression for rotation operator for spin $\dfrac12$. We see that the qubit tomogram does not contain Euler angle $\psi$ explicitly.

In \cite{Bell-type inequalities in classical probability theory} we found that the upper bound of Bell-type number within the framework of classical probability theory applied to the two-bit system is $2$. For the maximally entangled state
\begin{displaymath}
\hat{\rho}_{2\otimes2}=\left|\left| \begin {array}{cccc} 1/2&0&0&1/2\\\noalign{\medskip}0&0&0&0
\\\noalign{\medskip}0&0&0&0\\\noalign{\medskip}1/2&0&0&1/2\end {array}
\right|\right|
\eqno{\addtocounter{abcd}{1}(\arabic{abcd})}
\end{displaymath}
we observe violation of Bell-CHSH inequality for two-qubit system at a local point and near the upper bound as summarized in Table $1$. Following the methodology of tomographic probability theory we choose two directions $1,2$ and $1',2'$ for each of the qubits and predict quantum non-local correlations between their respective dichotomic variables. In our theory these non-local correlations rise naturally as an effect of stochastic nature of the matrices constructed of tomograms measured at various combinations of chosen qubit directions.

\begin{table}[h]
  \centering
\begin{tabular}{|c|c c|c c|c c|c c|}
  \hline
  \multirow{2}{*}{Angles}    & \multicolumn{4}{|c|}{Directions} & \multicolumn{4}{|c|}{Directions} \\
  \hline
            & 1 & 2 & 1' & 2' & 1 & 2 & 1' & 2' \\
  \hline
  $\varphi$ & $6.2800$ & $6.2800$  & $6.2800$ & $0.0416$ & $6.2639$ & $1.8010$  & $0.0190$ & $0.0200$\\
  \hline
  $\theta$  & $1.5834$ & $2.3705$  & $2.3705$ & $0.7962$ & $1.5834$ & $3.1410$  & $2.3620$ & $0.7910$\\
  \hline
           & \multicolumn{4}{|c|}{$B=2.4148$} & \multicolumn{4}{|c|}{$B=2.8284$}  \\
  \hline
\end{tabular}
\caption{Violation of Bell-CHSH inequality for qubit-qubit system near the Tsirelson bound.}
\end{table}
Taking the values of Euler angles from Table $1$ we are able to indicate the stochastic matrices resulting in the violation of Bell-CHSH inequality.
\begin{displaymath}
\hat{M}_{B=2.4148}=
\left|\left|
\begin{array}{cccc}
0.4264& 0.4263& 0.4999& 0.2490 \\
0.0735& 0.0736&0& 0.2509 \\
0.0735& 0.0736&0& 0.2509\\
0.4264& 0.4263& 0.4999&0.2490
\end{array}\right|\right|, \quad
\hat{M}_{B=2.8284}=\left|\left|
\begin{array}{cccc}
0.4280& 0.4255& 0.4278& 0.0742\\
0.0720& 0.0745&0.0722& 0.4258\\
0.0720& 0.0745&0.0722& 0.4258\\
0.4280& 0.4255& 0.4278&0.0742
\end{array} \right|\right|
\end{displaymath}

In Section $7.1$ of the Appendix the calculation of the explicit expression for $B_{2\otimes2}(\rho_{2\otimes2},\theta_1,\varphi_1,\theta_2,\varphi_2,\theta_3,\varphi_3,\theta_4,\varphi_4| \hat{C}=\hat{I}_{2\otimes2})$ are given.

The second column of directions in Table $1$ corresponds to the well known case of so called Tsirelson bound \cite{Tsirelson's theorem}. It manifests to the maximum entanglement in the qubit-qubit system in terms of the strongest quantum non-local correlations between the pair of qubits; a result which is beyond the scope of any classical local variable theory. The value of the upper bound is symmetric as it follows from the analytic expression of the Bell-CHSH form (see Section $7.1$ of Appendix). Numeric computation allows to obtain the lower bound as well:
\begin{displaymath}
\begin{array}{cccc}
\theta_1=1.9244 & \theta_3=1.9244 & \varphi_1=1.7351 & \varphi_3=1.7351 \\
\theta_2=0.5657 & \theta_4=0.5657 & \varphi_2=0.7780 & \varphi_4=0.7780 \\
\end{array}
\end{displaymath}
\begin{displaymath}
B=-2\sqrt{2}
\end{displaymath}
as expected the result is symmetric as it should be for any general linear correlation strength restriction.

In Fig. $1$ the results of calculations of Bell-type numbers as functions of the angle $\theta_1$ are shown. The picture graphically demonstrates Tsirelson's theorem \cite{Tsirelson's theorem}. In Section $7.1$ of the Appendix the surface of Bell-CHSH form is drawn near the Tsirelson bound of two correlating angles $\theta_2$ and $\theta_4$.

In the next several sections we study multiqudit states of higher dimensions ($d>2$) and calculate Bell-type numbers in classical probability theory. We then calculate Bell-type numbers and upper bounds of Bell-CHSH forms in the quantum tomographic theory and compare results. Our aim, as clarified in the Introduction is to point out examples of multiqudit states where Bell-type inequalities exist with various upper bounds analogous to Tsirelson bound in the lower dimension of $d=2$. With the implementation of quantum tomographic methods we see that there exist whole domains of parameters violating these Bell-type inequalities.

\begin{figure}[h!]
\includegraphics[width=500pt]{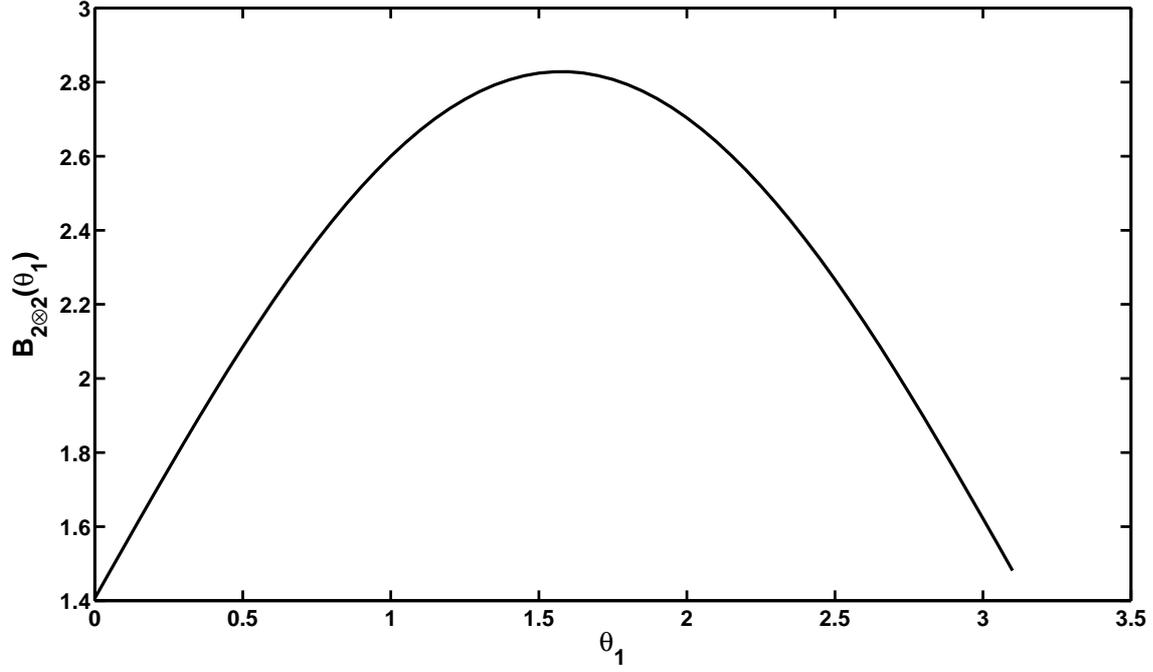}\\
\caption{The Bell-CHSH form $B_{2\otimes2}(\theta_1,\varphi_1,\theta_2,\varphi_2,\theta_3,\varphi_3,\theta_4,\varphi_4)$ in the vicinity of Tsirelson bound as a function of $\theta_1$.}
\end{figure}
Our approach allows to easily generalize the results obtained for the maximally entangled state $(8)$ over well-known Werner states for two qubits \cite{Werner states}. Indeed, for the state
\begin{displaymath}
\hat{\rho}^{W}(p)=\left|\left|
  \begin{array}{cccc}
    \dfrac{1+p}4 & 0 & 0 & \dfrac p2 \\
    0 & \dfrac{1-p}4 & 0 & 0 \\
    0 & 0 & \dfrac{1-p}4 & 0 \\
    \dfrac p2 & 0 & 0 & \dfrac{1+p}4 \\
  \end{array}
\right|\right|=\dfrac14\hat{1}+\dfrac p2\left|\left|
  \begin{array}{cccc}
    \dfrac12 & 0 & 0 & 1 \\
    0 & -\dfrac12 & 0 & 0 \\
    0 & 0 & -\dfrac12 & 0 \\
    1 & 0 & 0 & \dfrac12 \\
  \end{array}
\right|\right|
\end{displaymath}

The Bell-CHSH form assumes the expression:
\begin{gather*}
B^W_{2\otimes2}(\theta_1,\varphi_1,\theta_2,\varphi_2,\theta_3,\varphi_3,\theta_4,\varphi_4,p)=\\
=2\, p - 4\, p\, {\cos\!\left(\frac{\mathrm{{\theta}_1}}{2}\right)}^2 - 4\, p\, {\cos\!\left(\frac{\mathrm{{\theta}_3}}{2}\right)}^2 +4\, p\, {\cos\!\left(\frac{\mathrm{{\theta}_1}}{2}\right)}^2\, {\cos\!\left(\frac{\mathrm{{\theta}_3}}{2}\right)}^2 + \\
+4\, p\, {\cos\!\left(\frac{\mathrm{{\theta}_1}}{2}\right)}^2\, {\cos\!\left(\frac{\mathrm{{\theta}_4}}{2}\right)}^2 + 4\, p\, {\cos\!\left(\frac{\mathrm{{\theta}_2}}{2}\right)}^2\, {\cos\!\left(\frac{\mathrm{{\theta}_3}}{2}\right)}^2 - 4\, p\, {\cos\!\left(\frac{\mathrm{{\theta}_2}}{2}\right)}^2\, {\cos\!\left(\frac{\mathrm{{\theta}_4}}{2}\right)}^2 + \\
+4\, p\, \cos\!\left(\frac{\mathrm{{\theta}_1}}{2}\right)\, \cos\!\left(\frac{\mathrm{{\theta}_3}}{2}\right)\, \sin\!\left(\frac{\mathrm{{\theta}_1}}{2}\right)\, \sin\!\left(\frac{\mathrm{{\theta}_3}}{2}\right)\, \cos\!\left(\mathrm{{\varphi}_1}\right)\, \cos\!\left(\mathrm{{\varphi}_3}\right) + \\
+4\, p\, \cos\!\left(\frac{\mathrm{{\theta}_1}}{2}\right)\, \cos\!\left(\frac{\mathrm{{\theta}_4}}{2}\right)\, \sin\!\left(\frac{\mathrm{{\theta}_1}}{2}\right)\, \sin\!\left(\frac{\mathrm{{\theta}_4}}{2}\right)\, \cos\!\left(\mathrm{{\varphi}_1}\right)\, \cos\!\left(\mathrm{{\varphi}_4}\right) + \\
+4\, p\, \cos\!\left(\frac{\mathrm{{\theta}_2}}{2}\right)\, \cos\!\left(\frac{\mathrm{{\theta}_3}}{2}\right)\, \sin\!\left(\frac{\mathrm{{\theta}_2}}{2}\right)\, \sin\!\left(\frac{\mathrm{{\theta}_3}}{2}\right)\, \cos\!\left(\mathrm{{\varphi}_2}\right)\, \cos\!\left(\mathrm{{\varphi}_3}\right) - \\
-4\, p\, \cos\!\left(\frac{\mathrm{{\theta}_2}}{2}\right)\, \cos\!\left(\frac{\mathrm{{\theta}_4}}{2}\right)\, \sin\!\left(\frac{\mathrm{{\theta}_2}}{2}\right)\, \sin\!\left(\frac{\mathrm{{\theta}_4}}{2}\right)\, \cos\!\left(\mathrm{{\varphi}_2}\right)\, \cos\!\left(\mathrm{{\varphi}_4}\right) - \\
-4\, p\, \cos\!\left(\frac{\mathrm{{\theta}_1}}{2}\right)\, \cos\!\left(\frac{\mathrm{{\theta}_3}}{2}\right)\, \sin\!\left(\frac{\mathrm{{\theta}_1}}{2}\right)\, \sin\!\left(\frac{\mathrm{{\theta}_3}}{2}\right)\, \sin\!\left(\mathrm{{\varphi}_1}\right)\, \sin\!\left(\mathrm{{\varphi}_3}\right) -
\end{gather*}
\begin{gather*}
-4\, p\, \cos\!\left(\frac{\mathrm{{\theta}_1}}{2}\right)\, \cos\!\left(\frac{\mathrm{{\theta}_4}}{2}\right)\, \sin\!\left(\frac{\mathrm{{\theta}_1}}{2}\right)\, \sin\!\left(\frac{\mathrm{{\theta}_4}}{2}\right)\, \sin\!\left(\mathrm{{\varphi}_1}\right)\, \sin\!\left(\mathrm{{\varphi}_4}\right) - \\
-4\, p\, \cos\!\left(\frac{\mathrm{{\theta}_2}}{2}\right)\, \cos\!\left(\frac{\mathrm{{\theta}_3}}{2}\right)\, \sin\!\left(\frac{\mathrm{{\theta}_2}}{2}\right)\, \sin\!\left(\frac{\mathrm{{\theta}_3}}{2}\right)\, \sin\!\left(\mathrm{{\varphi}_2}\right)\, \sin\!\left(\mathrm{{\varphi}_3}\right) + \\
+4\, p\, \cos\!\left(\frac{\mathrm{{\theta}_2}}{2}\right)\, \cos\!\left(\frac{\mathrm{{\theta}_4}}{2}\right)\, \sin\!\left(\frac{\mathrm{{\theta}_2}}{2}\right)\, \sin\!\left(\frac{\mathrm{{\theta}_4}}{2}\right)\, \sin\!\left(\mathrm{{\varphi}_2}\right)\, \sin\!\left(\mathrm{{\varphi}_4}\right).
\end{gather*}
Figure $2$ shows the dependence of upper bounds of Bell-inequality for two qubits on the parameter $p$ that counts for the entanglement of the system. As it can be drawn from the graphic, the upper bound grows linearly in the interval of entanglement $\frac13<p<1$ reaching its maximum at the value of $p=1$.
\begin{center}
\begin{figure}[h!]
\includegraphics[width=500pt]{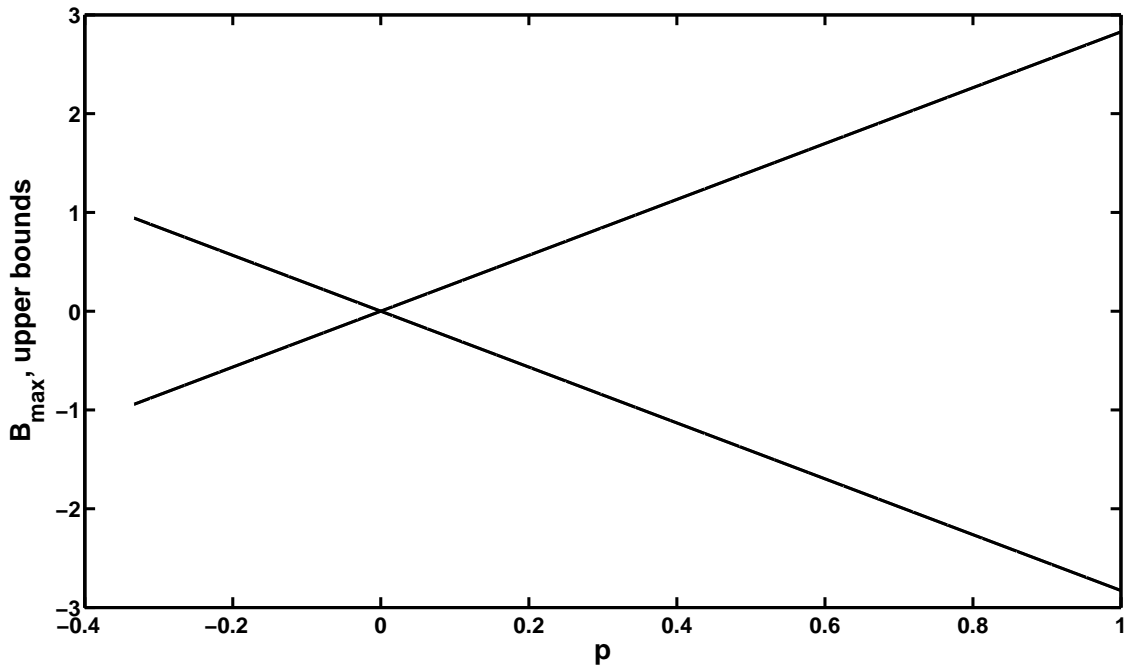}\\
\caption{The upper bounds of Werner states as function of parameter $p$.}
\end{figure}
\end{center}
The two-qubit system reviewed in this chapter provides an insight into studying higher dimension multiqudit states at least with respect to the amount of information that the Bell-CHSH form of a multipartite system stores in it. First of all the tomographic probability representation of multipartite states equips us with a universal tool both in classical and quantum realms. Procedure-wise the program of studying the system's Bell-type inequalities and upper bounds can be formulated the following way. Each subsystem contributes to the resulting joint-probability distribution of the composite system via a stochastic matrix. Information of correlations between various dynamic variables in the system (both in quantum and classical cases) are contained in the stochastic matrix of the composite state which can be either simply separable, separable or entangled (non-separable). This matrix enters a bilinear form named Bell-CHSH form through the average value of the product of correlating observables. We calculate all local (Bell-type numbers) and global maxima (the upper bound) of the Bell-CHSH bilinear form. We establish first the Bell-type numbers in classical probability theory for multi-level systems. Then we conduct the calculation of Bell-type numbers through quantum tomograms and compare results. This recipe allows one to detect possible Bell-type inequalities, classify them by their corresponding upper bounds and predict violation of these inequalities in quantum mechanics by observable angles rising in the measurement process of the quantum system. Moreover, given an entangled multiqudit state our method allows to find explicitly the domains in the parameter space where Bell-type inequalities are preserved and the small quantum domain where they are maximally violated. We will now realize this program for several interesting multiqudit states.

\section{Bell-CHSH type inequalities in qubit-qutrit system}
The correlations inside the qubit-qutrit system with two chosen directions $\mathbf{n}_1$ and $\mathbf{n}_2$ are given by the expectation value
\begin{displaymath}
\langle m_1m_2\rangle_{\mathbf{n}_1,\,\mathbf{n}_2}=\sum_{m_1=\pm1,m_2=\pm1,0} w\left(m_1,m_2,\mathbf{n}_1,\mathbf{n}_2\right)m_1m_2
\eqno{\addtocounter{abcd}{1}(\arabic{abcd})}
\end{displaymath}
 Just as in qubit systems the orientations of qudits in space are brought in by the necessity of measurements. In principle, one can choose any number of possible polarizations for each of the qudits in the composite system. However, since the signature of the correlation matrix of the bit-trit system can be presented in the form of a $6\times6$ matrix
\begin{displaymath}
\hat{\sigma}(2\otimes3)=\left|\left| \begin {array}{cccccc} 1&1&1&1&1&1\\\noalign{\medskip}0&0&0&0&0
&0\\\noalign{\medskip}-1&-1&-1&-1&-1&-1\\\noalign{\medskip}-1&-1&-1&-1
&-1&-1\\\noalign{\medskip}0&0&0&0&0&0\\\noalign{\medskip}1&1&1&1&1&1
\end {array} \right|\right|
\eqno{\addtocounter{abcd}{1}(\arabic{abcd})}
\end{displaymath}
we prefer ordering three directions of qutrit and two directions for qubit. This allows one to conduct maximal analysis of the resulting B-form. The expression for special sign matrix $(10)$ is obtained from formula $(9)$ by assuming all probabilities equal to $\dfrac16$. The special signature matrix $\hat{I}(2\otimes3)$ is received by transposing $\hat{\sigma}(2\otimes3)$ and inverting the sign of its last row\footnote{Generally speaking, various techniques of receiving various matrices $\hat{I}$ from the generic $\hat{\sigma}$ matrix are irrelevant to the problem of entanglement and subsequent violation of Bell-CHSH type inequalities. As long as the difference between the quantum and classical Bell-type numbers (upper bounds) are calculated via the same matrix $\hat{C}$ the developed approach stays perfectly legal.}.

To calculate the upper bounds of Bell-type inequalities i.e. obtain all possible Bell-type numbers,  we need to consider $2^8=256$ possible stochastic matrices related to $5$-possible directions, two per two-level system and three per the three-level system. However, not all of these vertices are physically feasible since the half of these vertices contain values of parameters $x_i$, $y_j$, $z_k$, $t_l$ $(i,j,k,l=1,2)$ prohibited by the requirement of positiveness of stochastic matrices i.e.
\begin{displaymath}
1-y_1-y_2\geq0, \qquad 1-z_1-z_2\geq0, \qquad 1-t_1-t_2\geq0.
\end{displaymath}
so the number of eligible vertices is reduced to $2^2\times3^3=108$. The two-level subsystem and the three-level subsystem are described by two stochastic matrices respectively
\begin{displaymath}
\hat{M}_\mathbf{x}=\left|\left|
  \begin{array}{cc}
    x_1  & x_2 \\
     1-x_1 & 1-x_2 \\
  \end{array}
\right|\right|,\qquad \hat{M}_{\mathbf{y},\mathbf{z},\mathbf{t}}=\left|\left|
  \begin{array}{ccc}
    y_1  & z_1 & t_1 \\
    y_2  & z_2 & t_2\\
    1-y_1-y_2 & 1-z_1-z_2 & 1-t_1-t_2 \\
  \end{array}
\right|\right|
\eqno{\addtocounter{abcd}{1}(\arabic{abcd})}
\end{displaymath}
where each of the variables $x_i$, $y_j$, $z_k$, $t_l$ $(i,j,k,l=1,2)$ assume values  between $0$ and $1$. The stochastic matrix of the composite system is $\hat{M}_{\mathbf{x},\mathbf{y},\mathbf{z},\mathbf{t}}=\hat{M}_\mathbf{x}\otimes\hat{M}_{\mathbf{y},\mathbf{z},\mathbf{t}}$ and the calculation for the first $27$ vertices for B-form $B\left(\mathbf{x},\mathbf{y},\mathbf{z},\mathbf{t}\right)=Tr\left(\hat{C}\hat{M}_{\mathbf{x},\mathbf{y},\mathbf{z},\mathbf{t}}\right)$ gives all-possible Bell-type numbers for the qubit-qutrit system in the general form as linear functions of the elements of the matrix $\hat{C}$. The list of Bell-type numbers in case of these vertices is given in the appendix.

The corresponding Bell-CHSH type numbers for a particular matrix
\begin{displaymath}
\hat{I}_{2\otimes3}=\left|\left|
\begin {array}{cccccc}
1&0&-1&-1&0&1\\\noalign{\medskip}
1&0&-1&-1&0&1\\\noalign{\medskip}
1&0&-1&-1&0&1\\\noalign{\medskip}
1&0&-1&-1&0&1\\\noalign{\medskip}
1&0&-1&-1&0&1\\\noalign{\medskip}
-1&0&1&1&0&-1
\end {array} \right|\right|
\eqno{\addtocounter{abcd}{1}(\arabic{abcd})}
\end{displaymath}
are calculated to be
\begin{displaymath}
B=-6,0,6,6,4,4,2,2,0,-4,0,2,6,8,12,6,0,2,2,4,4,6,0,2,0,0,6
\end{displaymath}
The values $\pm6$, $\pm8$ and $\pm12$ correspond to non-physical vertices (see formula $(1)$ in Section $7.2$ of Appendix for clarification) and should be discarded. Thus, we obtain
\begin{displaymath}
B^{classical}_{max}=4
\end{displaymath}
Consider now the quantum case. Let's examine the qubit-qutrit entangled state
\begin{displaymath}
\hat{\rho}_{2\otimes3}=\left|\left|
\begin{array}{cccccc}
1/2&0&0&0&0&1/2\\\noalign{\medskip}
0&0&0&0&0&0\\\noalign{\medskip}
0&0&0&0&0&0\\\noalign{\medskip}
0&0&0&0&0&0\\\noalign{\medskip}
0&0&0&0&0&0\\\noalign{\medskip}
1/2&0&0&0&0&1/2
\end{array} \right|\right|
\end{displaymath}

The classical bound $|B|=4$ can be reached on many points in the ten-dimensional phase space spanned by ten Euler angles, four for the two directions of qubit and six angles for the three directions of qutrit. One of the examples that can be instantly checked is:

\begin{table}[h]
  \centering
\begin{tabular}{|c|c c|c c c|c c|c c c|}
  \hline
  \multirow{2}{*}{Angles}    & \multicolumn{5}{|c|}{Directions} & \multicolumn{5}{|c|}{Directions} \\
  \hline
            & 1 & 2 & 1' & 2' & 3' & 1 & 2 & 1' & 2' & 3'\\
  \hline
  $\varphi$ & $0$ & $0$  & $0$   & $0$   & $0$   & $0.03$  & $3.19$ & $0$ & $0$ & $0$\\
  \hline
  $\theta$  & $0$ & $0$  & $\pi$ & $\pi$ & $\pi$ & $0.03$  & $0.22$ & $3.08$ & $3.09$ & $1.65$\\
  \hline
           & \multicolumn{5}{|c|}{$B=-4$} & \multicolumn{5}{|c|}{$B=4.0312$}  \\
  \hline
\end{tabular}
\caption{Violation of Bell-CHSH inequality for qubit-qutrit system}
\end{table}
The corresponding stochastic matrix easily calculates at Euler angles entering Table $2$ and gives:
\begin{displaymath}
\hat{M}_{B=-4}= \left|\left|
\begin{array}{cccccc}
0& 0& 0& 0& 0& 0 \\
\frac14& \frac14& \frac14& \frac14& \frac14& \frac14 \\\noalign{\medskip}
\frac12& \frac12& \frac12& \frac12& \frac12&\frac12 \\\noalign{\medskip}
\frac12& \frac12&\frac12& \frac12& \frac12& \frac12 \\\noalign{\medskip}
\frac14& \frac14& \frac14& \frac14& \frac14& \frac14 \\\noalign{\medskip}
0& 0& 0& 0& 0& 0
\end {array} \right|\right|
\end{displaymath}

The violating parameters for qubit-qutrit system were found through searching very small quantum domains. One example of such violation is presented by the second column of Table $2$. An example of a stronger violation of Bell-type inequality for qubit-qutrit system is given in the Table $6$ of Section $7.2$ of the Appendix.

\section{Bell-CHSH type inequalities in two-qutrit system}

The stochastic matrix $\hat{M}_{\mathbf{x},\mathbf{y},\mathbf{z}, \mathbf{r},\mathbf{s},\mathbf{t}}$ of the two three-level classical system reads:
\begin{displaymath}
\hat{M}_{\mathbf{x},\mathbf{y},\mathbf{z},\mathbf{r},\mathbf{s},\mathbf{t}}=\hat{M}_{\mathbf{x},\mathbf{y},\mathbf{z}}\otimes\hat{M}_{\mathbf{r},\mathbf{s},\mathbf{t}}
\end{displaymath}
where each of the matrices entering the final expression has the form of $(11)$. The explicit expressions for Bell-CHSH form and Bell-type numbers as functions of an arbitrary correlation matrix $C_{3\otimes3}$ can be found in Section $7.3$ of the Appendix.

There are $2^{12}=4096$ possible vertices and corresponding Bell-type numbers in this case. Again, because of the requirement of the positiveness of probabilities entering the stochastic matrices, this number is reduced to $3^6=729$ physically eligible vertices.

The upper bound corresponding to the particular case of the correlation matrix $\hat{I}_{3\otimes3}$ is found to be (see formula $(3)$ in Section $7.3$ of the Appendix)
\begin{displaymath}
B^{classical}_{max}=7
\end{displaymath}

\section{Bell-CHSH type inequalities in three-qubit system}
We consider now the system of three qubits. The classical probability theory approach treats the system as an eight-level system. The first $14$ Bell-type numbers as functions of arbitrary matrix elements $\hat{C}_{2\otimes2\otimes2}$ are presented in Section $7.4$ of the Appendix. Number of vertices i.e. number of possible Bell-type numbers in this case is $2^6=64$. Using the standard $\hat{I}_{2\otimes2\otimes2}$ matrix we obtain the following Bell-type numbers in classical tomographic probability theory:
\begin{displaymath}
B=0,\,\pm2,\,\pm4,\,\pm6
\end{displaymath}
The upper bound of Bell-type numbers is therefore
\begin{displaymath}
B^{classical}_{max}=6.
\end{displaymath}

In interesting discussion of three qubit system correlation functions was indicated in \cite{Explicit form of correlation function three-settings tight Bell inequalities for three qubits}. In Section $7.4$ of Appendix explicit expressions of classical and quantum correlation functions are presented in form of B-forms. These expressions change dramatically depending on the values of matrix elements of the sign matrix $\hat{C}$. By running a numeric search for local Bell-type inequality violations for several particular cases of $\hat{C}$ one needs to exceed the value $B=6$. The quantum violation of Bell-type inequalities here are very subtle due to large amount of possible inequalities not all of which can be violated.

\section{Conclusion}
To conclude, we formulate our main results. We studied several multiqudit systems in terms of their possible Bell-type numbers restricting correlations in multipartite states. Within the classical framework of probability theory we used simple tomographic approach to describe classical states as probability distribution vectors. In the quantum case the tomographic approach was sophisticated by entering quantum tomograms continuously depending on many Euler angles and strongly violating classical Bell-type inequalities. Bell-type numbers were constructed as maxima of some B-forms arising naturally in our tomographic description of states as average values of non-local correlation between subsystems. Our results are not complete as in case of three-qubit and two-qutrit systems the choice of the special matrix $\hat{C}$ becomes quite difficult and the violation of Bell-type inequalities becomes possible for many matrices. It is possible that there is no one universal matrix $\hat{I}$ for theThis matter needs a further analysis which we plan to present in our future works. The results of Bell-type numbers received in the article is presented in the Appendix with detailed description of numeric and analytic aspects of calculations. It is notable that the values of derivations of quantum mechanics from classical local variable theories based on probability description of both classical and quantum states depend on the same matrix $\hat{C}$ (or $\hat{I}$) i.e. there are hundreds of local variable theories violated by quantum mechanics.  Partial comparison of these theories are expressed in terms of Bell-numbers in classical local variable theories and in quantum mechanics. Some of our results include:
\begin{displaymath}
\begin{array}{ll}
B_{bit-bit}=2, & B_{bit-trit}=4 \\
B_{qubit-qubit}=2.8283, & B_{qubit-qutrit}=4.0612
\end{array}
\end{displaymath}
In our next work we plan to present the results for Bell-type numbers in two-qutrit and and three-qubit systems for several entangled states including the Werner family of states i.e. one-parametric families of Bell-type inequalities. Because the Bell-type numbers only partially characterize the phenomenon of entanglement in our next works we are going to introduce also various information entropies for multiqudit states hoping to establish (or prove otherwise) possible link between Bell-type numbers and entropies. In particular we are interested in domains where indications of entanglement are manifested both by maximal violation of Bell-type inequalities and relationships between entropies of subsystems with entropy of the composite system.
We would also like to aim at obtaining numeric and analytic results for a system of $N$ qudits within classical local variable theories to investigate further the origin of special matrix $\hat{C}$ strongly limiting possible Bell-type numbers. We intend to study in future works the multi-moment correlation measures, in comparison to those that consider only $2$nd moment (pairwise or quadratic studied in this paper) dependence. One of the interesting possibilities here is constructing matrix $\hat{C}$ through correlations like

\begin{displaymath}
C\propto\langle m^{n_1}_1m^{n_2}_2\dots m^{n_s}_s\rangle
\end{displaymath}

As it is well known, using multi-moment correlation functions it is possible to get a measure for more general dependence of Bell-type numbers on the initially given data (multi-variable states, correlation matrices, orientations in space etc.). We expect to increase the numeric "gap" between classical and the quantum Bell-type numbers.

In the present work the numeric calculations were carried out in MATLAB.

\section{Appendix}
In Appendix we have collected useful analytic and numeric results which demonstrate the calculation schemes for classical and quantum tomographic approaches. For convenience, Appendix is broken into subsections one per each of the systems studied.

\subsection{Two-qubit system}
The classical tomographic probability theory\cite{An introduction to the tomographic picture of quantum mechanics} applied to the system of two-bits provides us with Bell-type numbers as functions of elements of an arbitrary correlation matrix $\hat{C}$. The Bell-CHSH form for two qubits with a fixed matrix $\hat{I}_{2\otimes2}$ has the form
\begin{displaymath}
B_{2\otimes2}(x_1,x_2,y_1,y_2)=2+4\,x_{{1}}y_{{1}}+4\,x_{{1}}y_{{2}}-4\,x_{{1}}-4\,y_{{1}}-4\,x_{{2}}y_{{2}}+4\,x_{{2}}y_{{1}}.
\end{displaymath}
The results for $16$ Bell-type numbers in case of the familiar matrix $\hat{I}_{2\otimes2}$ are gathered in Table $3$. The combinations of vectors $\textbf{x}$ and $\textbf{y}$ represent the vertex of the polytope in the probability space of qubits. Bell-type numbers are the values of Bell-CHSH form taken at these vertices. The maximum value among these numbers, namely $B=2$ is the upper bound of Bell-type inequality for two-qubit system in classical probability theory.
\begin{table}[h]
  \centering
\begin{tabular}{|c|c|c||c|c|c||c|c|c||c|c|c|}
   \hline
   \textbf{x}  &  \textbf{y}  & B (\textbf{x},\textbf{y}) & \textbf{x}  &  \textbf{y}  & B (\textbf{x},\textbf{y}) & \textbf{x}  &  \textbf{y}  & B (\textbf{x},\textbf{y})
   & \textbf{x}  &  \textbf{y}  & B (\textbf{x},\textbf{y})\\
   \hline
   00 & 00 &  2 & 01 & 00 &  2 & 10 & 00 & -2 & 11 & 00 & -2 \\
   00 & 01 &  2 & 01 & 01 & -2 & 10 & 01 &  2 & 11 & 01 & -2 \\
   00 & 10 & -2 & 01 & 10 &  2 & 10 & 10 & -2 & 11 & 10 &  2 \\
   00 & 11 & -2 & 01 & 11 & -2 & 10 & 11 &  2 & 11 & 11 &  2 \\
  \hline
\end{tabular}
\caption{The Bell-type numbers of qubit-qubit system for the matrix $\hat{I}_{2\otimes2}$}
\end{table}

The matrix $\hat{P}$ reduces the dependence of $B\left(\mathbf{x},\mathbf{y},\mathbf{z},\mathbf{t}\right)$ on $12$ parameters in non-separable two-qubit state into four parameters as in separable state. It has the form:
\begin{displaymath}
\hat{P}=\left|\left| \begin {array}{rrrr} 1&-1&-1&-1\\\noalign{\medskip}1&-1&-1&-1
\\\noalign{\medskip}1&-1&-1&-1\\\noalign{\medskip}-1&1&1&1\end {array}
\right|\right|
\end{displaymath}
To calculate Bell-type numbers in quantum mechanics, we utilize two unitary matrices $\hat{U}_1$, $\hat{U}_2$ for the first qubit and two unitary matrices $\hat{V}_1$, $\hat{V}_2$ for the second qubit.

The Bell-CHSH form then calculates:
\begin{gather*}
B_{2\otimes2}(\theta_1,\varphi_1,\theta_2,\varphi_2,\theta_3,\varphi_3,\theta_4,\varphi_4) = \\
=2+4\, {\cos\!\left(\frac{\mathrm{{\theta}_1}}{2}\right)}^2\, {\cos\!\left(\frac{\mathrm{{\theta}_3}}{2}\right)}^2 +
4\, {\cos\!\left(\frac{\mathrm{{\theta}_1}}{2}\right)}^2\, {\cos\!\left(\frac{\mathrm{{\theta}_4}}{2}\right)}^2 + \\
+4\, {\cos\!\left(\frac{\mathrm{{\theta}_2}}{2}\right)}^2\, {\cos\!\left(\frac{\mathrm{{\theta}_3}}{2}\right)}^2 -
4\, {\cos\!\left(\frac{\mathrm{{\theta}_2}}{2}\right)}^2\, {\cos\!\left(\frac{\mathrm{{\theta}_4}}{2}\right)}^2 -
4\, {\cos\!\left(\frac{\mathrm{{\theta}_1}}{2}\right)}^2 - 4\, {\cos\!\left(\frac{\mathrm{{\theta}_3}}{2}\right)}^2 - \\
-4\, \cos\!\left(\frac{\mathrm{{\theta}_1}}{2}\right)\, \cos\!\left(\frac{\mathrm{{\theta}_3}}{2}\right)\, \sin\!\left(\frac{\mathrm{{\theta}_1}}{2}\right)\, \sin\!\left(\frac{\mathrm{{\theta}_3}}{2}\right)\, \cos\!\left(\mathrm{{\varphi}_1}\right)\, \cos\!\left(\mathrm{{\varphi}_3}\right) - \\
-4\, \cos\!\left(\frac{\mathrm{{\theta}_1}}{2}\right)\, \cos\!\left(\frac{\mathrm{{\theta}_4}}{2}\right)\, \sin\!\left(\frac{\mathrm{{\theta}_1}}{2}\right)\, \sin\!\left(\frac{\mathrm{{\theta}_4}}{2}\right)\, \cos\!\left(\mathrm{{\varphi}_1}\right)\, \cos\!\left(\mathrm{{\varphi}_4}\right) - \\
-4\, \cos\!\left(\frac{\mathrm{{\theta}_2}}{2}\right)\, \cos\!\left(\frac{\mathrm{{\theta}_3}}{2}\right)\, \sin\!\left(\frac{\mathrm{{\theta}_2}}{2}\right)\, \sin\!\left(\frac{\mathrm{{\theta}_3}}{2}\right)\, \cos\!\left(\mathrm{{\varphi}_2}\right)\, \cos\!\left(\mathrm{{\varphi}_3}\right) + \\
+4\, \cos\!\left(\frac{\mathrm{{\theta}_2}}{2}\right)\, \cos\!\left(\frac{\mathrm{{\theta}_4}}{2}\right)\, \sin\!\left(\frac{\mathrm{{\theta}_2}}{2}\right)\, \sin\!\left(\frac{\mathrm{{\theta}_4}}{2}\right)\, \cos\!\left(\mathrm{{\varphi}_2}\right)\, \cos\!\left(\mathrm{{\varphi}_4}\right) + \\
+4\, \cos\!\left(\frac{\mathrm{{\theta}_1}}{2}\right)\, \cos\!\left(\frac{\mathrm{{\theta}_3}}{2}\right)\, \sin\!\left(\frac{\mathrm{{\theta}_1}}{2}\right)\, \sin\!\left(\frac{\mathrm{{\theta}_3}}{2}\right)\, \sin\!\left(\mathrm{{\varphi}_1}\right)\, \sin\!\left(\mathrm{{\varphi}_3}\right) + \\
+4\, \cos\!\left(\frac{\mathrm{{\theta}_1}}{2}\right)\, \cos\!\left(\frac{\mathrm{{\theta}_4}}{2}\right)\, \sin\!\left(\frac{\mathrm{{\theta}_1}}{2}\right)\, \sin\!\left(\frac{\mathrm{{\theta}_4}}{2}\right)\, \sin\!\left(\mathrm{{\varphi}_1}\right)\, \sin\!\left(\mathrm{{\varphi}_4}\right) + \\
+4\, \cos\!\left(\frac{\mathrm{{\theta}_2}}{2}\right)\, \cos\!\left(\frac{\mathrm{{\theta}_3}}{2}\right)\, \sin\!\left(\frac{\mathrm{{\theta}_2}}{2}\right)\, \sin\!\left(\frac{\mathrm{{\theta}_3}}{2}\right)\, \sin\!\left(\mathrm{{\varphi}_2}\right)\, \sin\!\left(\mathrm{{\varphi}_3}\right) - \\
-4\, \cos\!\left(\frac{\mathrm{{\theta}_2}}{2}\right)\, \cos\!\left(\frac{\mathrm{{\theta}_4}}{2}\right)\, \sin\!\left(\frac{\mathrm{{\theta}_2}}{2}\right)\, \sin\!\left(\frac{\mathrm{{\theta}_4}}{2}\right)\, \sin\!\left(\mathrm{{\varphi}_2}\right)\, \sin\!\left(\mathrm{{\varphi}_4}\right)
\end{gather*}

In Fig. $3$ the Bell-CHSH form $B_{2\otimes2}(\theta_1,\varphi_1,\theta_2,\varphi_2,\theta_3,\varphi_3,\theta_4,\varphi_4)$ as function of $\theta_2$ and $\theta_4$ was plotted. The values of the rest of the Euler angles were taken very close to those which give Tsirelson bound (see Table $1$).
\begin{figure}[h!]
\includegraphics[width=500pt]{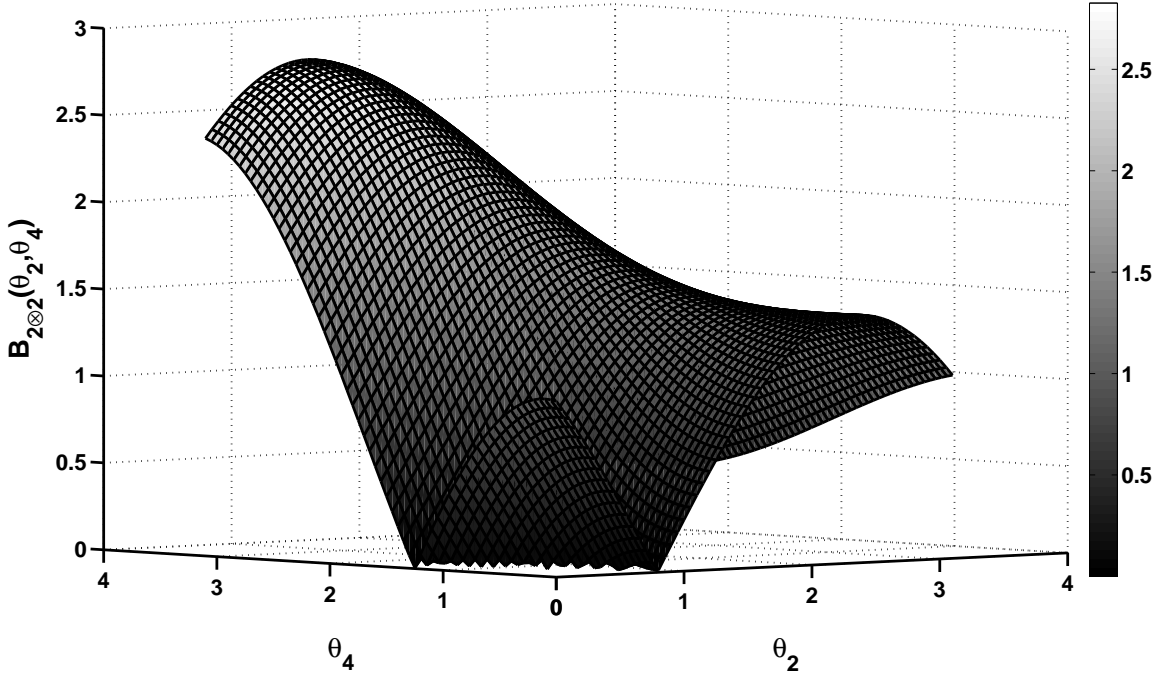}\\
\caption{The Bell-CHSH form $B_{2\otimes2}(\theta_1,\varphi_1,\theta_2,\varphi_2,\theta_3,\varphi_3,\theta_4,\varphi_4)$ as a function of $\theta_2$ and $\theta_4$}
\end{figure}The white area on the plot indicates the small quantum domain where the Bell-CHSH form exceeds the classical restraint of $B=2$. The global maxima of the function is the Tsirelson bound.
The Bell-CHSH form in case of a general matrix $\hat{C}$ reads:
\begin{gather*}
B = \frac{\mathrm{c_{11}}}{2} + \frac{\mathrm{c_{14}}}{2} + \frac{\mathrm{c_{21}}}{2} + \frac{\mathrm{c_{24}}}{2} + \frac{\mathrm{c_{31}}}{2} + \frac{\mathrm{c_{34}}}{2} + \frac{\mathrm{c_{41}}}{2} + \frac{\mathrm{c_{44}}}{2} - \\
\frac{\mathrm{c_{11}}\, {\cos\!\left(\frac{\mathrm{{\theta}_1}}{2}\right)}^2}{2} + \frac{\mathrm{c_{12}}\, {\cos\!\left(\frac{\mathrm{{\theta}_1}}{2}\right)}^2}{2} - \frac{\mathrm{c_{11}}\, {\cos\!\left(\frac{\mathrm{{\theta}_3}}{2}\right)}^2}{2} + \frac{\mathrm{c_{13}}\, {\cos\!\left(\frac{\mathrm{{\theta}_1}}{2}\right)}^2}{2} + \frac{\mathrm{c_{12}}\, {\cos\!\left(\frac{\mathrm{{\theta}_3}}{2}\right)}^2}{2} - \frac{\mathrm{c_{14}}\, {\cos\!\left(\frac{\mathrm{{\theta}_1}}{2}\right)}^2}{2} + \\
\frac{\mathrm{c_{13}}\, {\cos\!\left(\frac{\mathrm{{\theta}_3}}{2}\right)}^2}{2} - \frac{\mathrm{c_{14}}\, {\cos\!\left(\frac{\mathrm{{\theta}_3}}{2}\right)}^2}{2} - \frac{\mathrm{c_{21}}\, {\cos\!\left(\frac{\mathrm{{\theta}_1}}{2}\right)}^2}{2} + \frac{\mathrm{c_{22}}\, {\cos\!\left(\frac{\mathrm{{\theta}_1}}{2}\right)}^2}{2} + \frac{\mathrm{c_{23}}\, {\cos\!\left(\frac{\mathrm{{\theta}_1}}{2}\right)}^2}{2} - \frac{\mathrm{c_{21}}\, {\cos\!\left(\frac{\mathrm{{\theta}_4}}{2}\right)}^2}{2} - \frac{\mathrm{c_{24}}\, {\cos\!\left(\frac{\mathrm{{\theta}_1}}{2}\right)}^2}{2} + \\
\frac{\mathrm{c_{22}}\, {\cos\!\left(\frac{\mathrm{{\theta}_4}}{2}\right)}^2}{2} + \frac{\mathrm{c_{23}}\, {\cos\!\left(\frac{\mathrm{{\theta}_4}}{2}\right)}^2}{2} - \frac{\mathrm{c_{24}}\, {\cos\!\left(\frac{\mathrm{{\theta}_4}}{2}\right)}^2}{2} - \frac{\mathrm{c_{31}}\, {\cos\!\left(\frac{\mathrm{{\theta}_2}}{2}\right)}^2}{2} - \frac{\mathrm{c_{31}}\, {\cos\!\left(\frac{\mathrm{{\theta}_3}}{2}\right)}^2}{2} + \frac{\mathrm{c_{32}}\, {\cos\!\left(\frac{\mathrm{{\theta}_2}}{2}\right)}^2}{2} + \frac{\mathrm{c_{32}}\, {\cos\!\left(\frac{\mathrm{{\theta}_3}}{2}\right)}^2}{2} + \\
\frac{\mathrm{c_{33}}\, {\cos\!\left(\frac{\mathrm{{\theta}_2}}{2}\right)}^2}{2} + \frac{\mathrm{c_{33}}\, {\cos\!\left(\frac{\mathrm{{\theta}_3}}{2}\right)}^2}{2} - \frac{\mathrm{c_{34}}\, {\cos\!\left(\frac{\mathrm{{\theta}_2}}{2}\right)}^2}{2} - \frac{\mathrm{c_{34}}\, {\cos\!\left(\frac{\mathrm{{\theta}_3}}{2}\right)}^2}{2} - \frac{\mathrm{c_{41}}\, {\cos\!\left(\frac{\mathrm{{\theta}_2}}{2}\right)}^2}{2} + \frac{\mathrm{c_{42}}\, {\cos\!\left(\frac{\mathrm{{\theta}_2}}{2}\right)}^2}{2} - \frac{\mathrm{c_{41}}\, {\cos\!\left(\frac{\mathrm{{\theta}_4}}{2}\right)}^2}{2} + \\
\frac{\mathrm{c_{43}}\, {\cos\!\left(\frac{\mathrm{{\theta}_2}}{2}\right)}^2}{2} + \frac{\mathrm{c_{42}}\, {\cos\!\left(\frac{\mathrm{{\theta}_4}}{2}\right)}^2}{2} - \frac{\mathrm{c_{44}}\, {\cos\!\left(\frac{\mathrm{{\theta}_2}}{2}\right)}^2}{2} + \frac{\mathrm{c_{43}}\, {\cos\!\left(\frac{\mathrm{{\theta}_4}}{2}\right)}^2}{2} - \frac{\mathrm{c_{44}}\, {\cos\!\left(\frac{\mathrm{{\theta}_4}}{2}\right)}^2}{2} + \\
\mathrm{c_{11}}\, {\cos\!\left(\frac{\mathrm{{\theta}_1}}{2}\right)}^2\, {\cos\!\left(\frac{\mathrm{{\theta}_3}}{2}\right)}^2 - \mathrm{c_{12}}\, {\cos\!\left(\frac{\mathrm{{\theta}_1}}{2}\right)}^2\, {\cos\!\left(\frac{\mathrm{{\theta}_3}}{2}\right)}^2 - \mathrm{c_{13}}\, {\cos\!\left(\frac{\mathrm{{\theta}_1}}{2}\right)}^2\, {\cos\!\left(\frac{\mathrm{{\theta}_3}}{2}\right)}^2 + \mathrm{c_{14}}\, {\cos\!\left(\frac{\mathrm{{\theta}_1}}{2}\right)}^2\, {\cos\!\left(\frac{\mathrm{{\theta}_3}}{2}\right)}^2 + \\
\mathrm{c_{21}}\, {\cos\!\left(\frac{\mathrm{{\theta}_1}}{2}\right)}^2\, {\cos\!\left(\frac{\mathrm{{\theta}_4}}{2}\right)}^2 - \mathrm{c_{22}}\, {\cos\!\left(\frac{\mathrm{{\theta}_1}}{2}\right)}^2\, {\cos\!\left(\frac{\mathrm{{\theta}_4}}{2}\right)}^2 - \mathrm{c_{23}}\, {\cos\!\left(\frac{\mathrm{{\theta}_1}}{2}\right)}^2\, {\cos\!\left(\frac{\mathrm{{\theta}_4}}{2}\right)}^2 + \mathrm{c_{24}}\, {\cos\!\left(\frac{\mathrm{{\theta}_1}}{2}\right)}^2\, {\cos\!\left(\frac{\mathrm{{\theta}_4}}{2}\right)}^2 + \\
\mathrm{c_{31}}\, {\cos\!\left(\frac{\mathrm{{\theta}_2}}{2}\right)}^2\, {\cos\!\left(\frac{\mathrm{{\theta}_3}}{2}\right)}^2 - \mathrm{c_{32}}\, {\cos\!\left(\frac{\mathrm{{\theta}_2}}{2}\right)}^2\, {\cos\!\left(\frac{\mathrm{{\theta}_3}}{2}\right)}^2 - \mathrm{c_{33}}\, {\cos\!\left(\frac{\mathrm{{\theta}_2}}{2}\right)}^2\, {\cos\!\left(\frac{\mathrm{{\theta}_3}}{2}\right)}^2 + \mathrm{c_{34}}\, {\cos\!\left(\frac{\mathrm{{\theta}_2}}{2}\right)}^2\, {\cos\!\left(\frac{\mathrm{{\theta}_3}}{2}\right)}^2 + \\
\mathrm{c_{41}}\, {\cos\!\left(\frac{\mathrm{{\theta}_2}}{2}\right)}^2\, {\cos\!\left(\frac{\mathrm{{\theta}_4}}{2}\right)}^2 - \mathrm{c_{42}}\, {\cos\!\left(\frac{\mathrm{{\theta}_2}}{2}\right)}^2\, {\cos\!\left(\frac{\mathrm{{\theta}_4}}{2}\right)}^2 - \mathrm{c_{43}}\, {\cos\!\left(\frac{\mathrm{{\theta}_2}}{2}\right)}^2\, {\cos\!\left(\frac{\mathrm{{\theta}_4}}{2}\right)}^2 + \mathrm{c_{44}}\, {\cos\!\left(\frac{\mathrm{{\theta}_2}}{2}\right)}^2\, {\cos\!\left(\frac{\mathrm{{\theta}_4}}{2}\right)}^2 - \\
\mathrm{c_{11}}\, \cos\!\left(\frac{\mathrm{{\theta}_1}}{2}\right)\, \cos\!\left(\frac{\mathrm{{\theta}_3}}{2}\right)\, \sin\!\left(\frac{\mathrm{{\theta}_1}}{2}\right)\, \sin\!\left(\frac{\mathrm{{\theta}_3}}{2}\right)\, \cos\!\left(\mathrm{{\varphi}_1}\right)\, \cos\!\left(\mathrm{{\varphi}_3}\right) + \\
\mathrm{c_{12}}\, \cos\!\left(\frac{\mathrm{{\theta}_1}}{2}\right)\, \cos\!\left(\frac{\mathrm{{\theta}_3}}{2}\right)\, \sin\!\left(\frac{\mathrm{{\theta}_1}}{2}\right)\, \sin\!\left(\frac{\mathrm{{\theta}_3}}{2}\right)\, \cos\!\left(\mathrm{{\varphi}_1}\right)\, \cos\!\left(\mathrm{{\varphi}_3}\right) + \\
\end{gather*}
\begin{gather*}
\mathrm{c_{13}}\, \cos\!\left(\frac{\mathrm{{\theta}_1}}{2}\right)\, \cos\!\left(\frac{\mathrm{{\theta}_3}}{2}\right)\, \sin\!\left(\frac{\mathrm{{\theta}_1}}{2}\right)\, \sin\!\left(\frac{\mathrm{{\theta}_3}}{2}\right)\, \cos\!\left(\mathrm{{\varphi}_1}\right)\, \cos\!\left(\mathrm{{\varphi}_3}\right) - \\
\mathrm{c_{14}}\, \cos\!\left(\frac{\mathrm{{\theta}_1}}{2}\right)\, \cos\!\left(\frac{\mathrm{{\theta}_3}}{2}\right)\, \sin\!\left(\frac{\mathrm{{\theta}_1}}{2}\right)\, \sin\!\left(\frac{\mathrm{{\theta}_3}}{2}\right)\, \cos\!\left(\mathrm{{\varphi}_1}\right)\, \cos\!\left(\mathrm{{\varphi}_3}\right) - \\
\mathrm{c_{21}}\, \cos\!\left(\frac{\mathrm{{\theta}_1}}{2}\right)\, \cos\!\left(\frac{\mathrm{{\theta}_4}}{2}\right)\, \sin\!\left(\frac{\mathrm{{\theta}_1}}{2}\right)\, \sin\!\left(\frac{\mathrm{{\theta}_4}}{2}\right)\, \cos\!\left(\mathrm{{\varphi}_1}\right)\, \cos\!\left(\mathrm{{\varphi}_4}\right) + \\
\mathrm{c_{22}}\, \cos\!\left(\frac{\mathrm{{\theta}_1}}{2}\right)\, \cos\!\left(\frac{\mathrm{{\theta}_4}}{2}\right)\, \sin\!\left(\frac{\mathrm{{\theta}_1}}{2}\right)\, \sin\!\left(\frac{\mathrm{{\theta}_4}}{2}\right)\, \cos\!\left(\mathrm{{\varphi}_1}\right)\, \cos\!\left(\mathrm{{\varphi}_4}\right) + \\
\mathrm{c_{23}}\, \cos\!\left(\frac{\mathrm{{\theta}_1}}{2}\right)\, \cos\!\left(\frac{\mathrm{{\theta}_4}}{2}\right)\, \sin\!\left(\frac{\mathrm{{\theta}_1}}{2}\right)\, \sin\!\left(\frac{\mathrm{{\theta}_4}}{2}\right)\, \cos\!\left(\mathrm{{\varphi}_1}\right)\, \cos\!\left(\mathrm{{\varphi}_4}\right) - \\
\mathrm{c_{24}}\, \cos\!\left(\frac{\mathrm{{\theta}_1}}{2}\right)\, \cos\!\left(\frac{\mathrm{{\theta}_4}}{2}\right)\, \sin\!\left(\frac{\mathrm{{\theta}_1}}{2}\right)\, \sin\!\left(\frac{\mathrm{{\theta}_4}}{2}\right)\, \cos\!\left(\mathrm{{\varphi}_1}\right)\, \cos\!\left(\mathrm{{\varphi}_4}\right) - \\
\mathrm{c_{31}}\, \cos\!\left(\frac{\mathrm{{\theta}_2}}{2}\right)\, \cos\!\left(\frac{\mathrm{{\theta}_3}}{2}\right)\, \sin\!\left(\frac{\mathrm{{\theta}_2}}{2}\right)\, \sin\!\left(\frac{\mathrm{{\theta}_3}}{2}\right)\, \cos\!\left(\mathrm{{\varphi}_2}\right)\, \cos\!\left(\mathrm{{\varphi}_3}\right) + \\
\mathrm{c_{32}}\, \cos\!\left(\frac{\mathrm{{\theta}_2}}{2}\right)\, \cos\!\left(\frac{\mathrm{{\theta}_3}}{2}\right)\, \sin\!\left(\frac{\mathrm{{\theta}_2}}{2}\right)\, \sin\!\left(\frac{\mathrm{{\theta}_3}}{2}\right)\, \cos\!\left(\mathrm{{\varphi}_2}\right)\, \cos\!\left(\mathrm{{\varphi}_3}\right) + \\
\mathrm{c_{33}}\, \cos\!\left(\frac{\mathrm{{\theta}_2}}{2}\right)\, \cos\!\left(\frac{\mathrm{{\theta}_3}}{2}\right)\, \sin\!\left(\frac{\mathrm{{\theta}_2}}{2}\right)\, \sin\!\left(\frac{\mathrm{{\theta}_3}}{2}\right)\, \cos\!\left(\mathrm{{\varphi}_2}\right)\, \cos\!\left(\mathrm{{\varphi}_3}\right) - \\
\mathrm{c_{34}}\, \cos\!\left(\frac{\mathrm{{\theta}_2}}{2}\right)\, \cos\!\left(\frac{\mathrm{{\theta}_3}}{2}\right)\, \sin\!\left(\frac{\mathrm{{\theta}_2}}{2}\right)\, \sin\!\left(\frac{\mathrm{{\theta}_3}}{2}\right)\, \cos\!\left(\mathrm{{\varphi}_2}\right)\, \cos\!\left(\mathrm{{\varphi}_3}\right) - \\
\mathrm{c_{41}}\, \cos\!\left(\frac{\mathrm{{\theta}_2}}{2}\right)\, \cos\!\left(\frac{\mathrm{{\theta}_4}}{2}\right)\, \sin\!\left(\frac{\mathrm{{\theta}_2}}{2}\right)\, \sin\!\left(\frac{\mathrm{{\theta}_4}}{2}\right)\, \cos\!\left(\mathrm{{\varphi}_2}\right)\, \cos\!\left(\mathrm{{\varphi}_4}\right) + \\
\mathrm{c_{42}}\, \cos\!\left(\frac{\mathrm{{\theta}_2}}{2}\right)\, \cos\!\left(\frac{\mathrm{{\theta}_4}}{2}\right)\, \sin\!\left(\frac{\mathrm{{\theta}_2}}{2}\right)\, \sin\!\left(\frac{\mathrm{{\theta}_4}}{2}\right)\, \cos\!\left(\mathrm{{\varphi}_2}\right)\, \cos\!\left(\mathrm{{\varphi}_4}\right) + \\
\mathrm{c_{43}}\, \cos\!\left(\frac{\mathrm{{\theta}_2}}{2}\right)\, \cos\!\left(\frac{\mathrm{{\theta}_4}}{2}\right)\, \sin\!\left(\frac{\mathrm{{\theta}_2}}{2}\right)\, \sin\!\left(\frac{\mathrm{{\theta}_4}}{2}\right)\, \cos\!\left(\mathrm{{\varphi}_2}\right)\, \cos\!\left(\mathrm{{\varphi}_4}\right) - \\
\mathrm{c_{44}}\, \cos\!\left(\frac{\mathrm{{\theta}_2}}{2}\right)\, \cos\!\left(\frac{\mathrm{{\theta}_4}}{2}\right)\, \sin\!\left(\frac{\mathrm{{\theta}_2}}{2}\right)\, \sin\!\left(\frac{\mathrm{{\theta}_4}}{2}\right)\, \cos\!\left(\mathrm{{\varphi}_2}\right)\, \cos\!\left(\mathrm{{\varphi}_4}\right) + \\
\mathrm{c_{11}}\, \cos\!\left(\frac{\mathrm{{\theta}_1}}{2}\right)\, \cos\!\left(\frac{\mathrm{{\theta}_3}}{2}\right)\, \sin\!\left(\frac{\mathrm{{\theta}_1}}{2}\right)\, \sin\!\left(\frac{\mathrm{{\theta}_3}}{2}\right)\, \sin\!\left(\mathrm{{\varphi}_1}\right)\, \sin\!\left(\mathrm{{\varphi}_3}\right) - \\
\mathrm{c_{12}}\, \cos\!\left(\frac{\mathrm{{\theta}_1}}{2}\right)\, \cos\!\left(\frac{\mathrm{{\theta}_3}}{2}\right)\, \sin\!\left(\frac{\mathrm{{\theta}_1}}{2}\right)\, \sin\!\left(\frac{\mathrm{{\theta}_3}}{2}\right)\, \sin\!\left(\mathrm{{\varphi}_1}\right)\, \sin\!\left(\mathrm{{\varphi}_3}\right) - \\
\mathrm{c_{13}}\, \cos\!\left(\frac{\mathrm{{\theta}_1}}{2}\right)\, \cos\!\left(\frac{\mathrm{{\theta}_3}}{2}\right)\, \sin\!\left(\frac{\mathrm{{\theta}_1}}{2}\right)\, \sin\!\left(\frac{\mathrm{{\theta}_3}}{2}\right)\, \sin\!\left(\mathrm{{\varphi}_1}\right)\, \sin\!\left(\mathrm{{\varphi}_3}\right) + \\
\mathrm{c_{14}}\, \cos\!\left(\frac{\mathrm{{\theta}_1}}{2}\right)\, \cos\!\left(\frac{\mathrm{{\theta}_3}}{2}\right)\, \sin\!\left(\frac{\mathrm{{\theta}_1}}{2}\right)\, \sin\!\left(\frac{\mathrm{{\theta}_3}}{2}\right)\, \sin\!\left(\mathrm{{\varphi}_1}\right)\, \sin\!\left(\mathrm{{\varphi}_3}\right) + \\
\mathrm{c_{21}}\, \cos\!\left(\frac{\mathrm{{\theta}_1}}{2}\right)\, \cos\!\left(\frac{\mathrm{{\theta}_4}}{2}\right)\, \sin\!\left(\frac{\mathrm{{\theta}_1}}{2}\right)\, \sin\!\left(\frac{\mathrm{{\theta}_4}}{2}\right)\, \sin\!\left(\mathrm{{\varphi}_1}\right)\, \sin\!\left(\mathrm{{\varphi}_4}\right) - \\
\mathrm{c_{22}}\, \cos\!\left(\frac{\mathrm{{\theta}_1}}{2}\right)\, \cos\!\left(\frac{\mathrm{{\theta}_4}}{2}\right)\, \sin\!\left(\frac{\mathrm{{\theta}_1}}{2}\right)\, \sin\!\left(\frac{\mathrm{{\theta}_4}}{2}\right)\, \sin\!\left(\mathrm{{\varphi}_1}\right)\, \sin\!\left(\mathrm{{\varphi}_4}\right) - \\
\mathrm{c_{23}}\, \cos\!\left(\frac{\mathrm{{\theta}_1}}{2}\right)\, \cos\!\left(\frac{\mathrm{{\theta}_4}}{2}\right)\, \sin\!\left(\frac{\mathrm{{\theta}_1}}{2}\right)\, \sin\!\left(\frac{\mathrm{{\theta}_4}}{2}\right)\, \sin\!\left(\mathrm{{\varphi}_1}\right)\, \sin\!\left(\mathrm{{\varphi}_4}\right) + \\
\mathrm{c_{24}}\, \cos\!\left(\frac{\mathrm{{\theta}_1}}{2}\right)\, \cos\!\left(\frac{\mathrm{{\theta}_4}}{2}\right)\, \sin\!\left(\frac{\mathrm{{\theta}_1}}{2}\right)\, \sin\!\left(\frac{\mathrm{{\theta}_4}}{2}\right)\, \sin\!\left(\mathrm{{\varphi}_1}\right)\, \sin\!\left(\mathrm{{\varphi}_4}\right) + \\
\mathrm{c_{31}}\, \cos\!\left(\frac{\mathrm{{\theta}_2}}{2}\right)\, \cos\!\left(\frac{\mathrm{{\theta}_3}}{2}\right)\, \sin\!\left(\frac{\mathrm{{\theta}_2}}{2}\right)\, \sin\!\left(\frac{\mathrm{{\theta}_3}}{2}\right)\, \sin\!\left(\mathrm{{\varphi}_2}\right)\, \sin\!\left(\mathrm{{\varphi}_3}\right) - \\
\mathrm{c_{32}}\, \cos\!\left(\frac{\mathrm{{\theta}_2}}{2}\right)\, \cos\!\left(\frac{\mathrm{{\theta}_3}}{2}\right)\, \sin\!\left(\frac{\mathrm{{\theta}_2}}{2}\right)\, \sin\!\left(\frac{\mathrm{{\theta}_3}}{2}\right)\, \sin\!\left(\mathrm{{\varphi}_2}\right)\, \sin\!\left(\mathrm{{\varphi}_3}\right) - \\
\mathrm{c_{33}}\, \cos\!\left(\frac{\mathrm{{\theta}_2}}{2}\right)\, \cos\!\left(\frac{\mathrm{{\theta}_3}}{2}\right)\, \sin\!\left(\frac{\mathrm{{\theta}_2}}{2}\right)\, \sin\!\left(\frac{\mathrm{{\theta}_3}}{2}\right)\, \sin\!\left(\mathrm{{\varphi}_2}\right)\, \sin\!\left(\mathrm{{\varphi}_3}\right) + \\
\mathrm{c_{34}}\, \cos\!\left(\frac{\mathrm{{\theta}_2}}{2}\right)\, \cos\!\left(\frac{\mathrm{{\theta}_3}}{2}\right)\, \sin\!\left(\frac{\mathrm{{\theta}_2}}{2}\right)\, \sin\!\left(\frac{\mathrm{{\theta}_3}}{2}\right)\, \sin\!\left(\mathrm{{\varphi}_2}\right)\, \sin\!\left(\mathrm{{\varphi}_3}\right) + \\
\end{gather*}
\begin{gather*}
\mathrm{c_{41}}\, \cos\!\left(\frac{\mathrm{{\theta}_2}}{2}\right)\, \cos\!\left(\frac{\mathrm{{\theta}_4}}{2}\right)\, \sin\!\left(\frac{\mathrm{{\theta}_2}}{2}\right)\, \sin\!\left(\frac{\mathrm{{\theta}_4}}{2}\right)\, \sin\!\left(\mathrm{{\varphi}_2}\right)\, \sin\!\left(\mathrm{{\varphi}_4}\right) - \\
\mathrm{c_{42}}\, \cos\!\left(\frac{\mathrm{{\theta}_2}}{2}\right)\, \cos\!\left(\frac{\mathrm{{\theta}_4}}{2}\right)\, \sin\!\left(\frac{\mathrm{{\theta}_2}}{2}\right)\, \sin\!\left(\frac{\mathrm{{\theta}_4}}{2}\right)\, \sin\!\left(\mathrm{{\varphi}_2}\right)\, \sin\!\left(\mathrm{{\varphi}_4}\right) - \\
\mathrm{c_{43}}\, \cos\!\left(\frac{\mathrm{{\theta}_2}}{2}\right)\, \cos\!\left(\frac{\mathrm{{\theta}_4}}{2}\right)\, \sin\!\left(\frac{\mathrm{{\theta}_2}}{2}\right)\, \sin\!\left(\frac{\mathrm{{\theta}_4}}{2}\right)\, \sin\!\left(\mathrm{{\varphi}_2}\right)\, \sin\!\left(\mathrm{{\varphi}_4}\right) + \\
\mathrm{c_{44}}\, \cos\!\left(\frac{\mathrm{{\theta}_2}}{2}\right)\, \cos\!\left(\frac{\mathrm{{\theta}_4}}{2}\right)\, \sin\!\left(\frac{\mathrm{{\theta}_2}}{2}\right)\, \sin\!\left(\frac{\mathrm{{\theta}_4}}{2}\right)\, \sin\!\left(\mathrm{{\varphi}_2}\right)\, \sin\!\left(\mathrm{{\varphi}_4}\right)
\end{gather*}

For illustration of the situation when Bell-type inequality is preserved by the two-qubit system we present the following values of Bell-CHSH forms gathered in the Table $4$. The table shows values of the Bell-CHSH form corresponding to two possible orientations for each of the qubits denoted $1,2$ and $1',2'$ accordingly.

\begin{table}[h]
  \centering
\begin{tabular}{|c|c c|c c|c c|c c|}
  \hline
  \multirow{2}{*}{Angles}    & \multicolumn{4}{|c|}{Directions} & \multicolumn{4}{|c|}{Directions}  \\
  \hline
            & 1 & 2 & 1' & 2' & 1 & 2 & 1' & 2' \\
  \hline
  $\varphi$ & $0$ & $\dfrac{\pi}3$  & $\pi$   & $0$   & $\dfrac{\pi}4$   & $\dfrac{\pi}4$  & $0$ & $\dfrac{\pi}4$ \\
  \hline
  $\theta$  & $\dfrac{\pi}2$ & $0$  & $0$ & $\dfrac{\pi}4$ & $\dfrac{\pi}6$ & $\dfrac{\pi}4$  & $\dfrac{\pi}4$ & $\dfrac{\pi}4$ \\
  \hline
           & \multicolumn{4}{|c|}{$B=1$} & \multicolumn{4}{|c|}{$B\approx1.83$}  \\
  \hline
\end{tabular}
\caption{Bell-CHSH inequality validation for qubit-qubit system}
\end{table}
Numeric computation also allows to recover the stochastic matrices corresponding to these values of Bell-CHSH form. For the value of $B=1$ the stochastic matrix reads:
\begin{displaymath}
\hat{M}_{B=1}=\left|\left| \begin {array}{cccc} 1/4&1/4+1/8\,\sqrt {2}&1/2&1/4+1/8\,
\sqrt {2}\\\noalign{\medskip}1/4&1/4-1/8\,\sqrt {2}&0&1/4-1/8\,\sqrt {
2}\\\noalign{\medskip}1/4&1/4-1/8\,\sqrt {2}&0&1/4-1/8\,\sqrt {2}
\\\noalign{\medskip}1/4&1/4+1/8\,\sqrt {2}&1/2&1/4+1/8\,\sqrt {2}
\end {array}\right|\right|,
\end{displaymath}
For an increased but still "classically suppressed" value of $B\approx1.83$ its stochastic matrix reads:
\begin{displaymath}
\hat{M}_{B\approx1.83}= \left|\left|
\begin {array}{cccc}
5/16+1/16\,\sqrt {6}&1/4+1/16 \,\sqrt {6}&3/8+1/16\,\sqrt {2}&3/8\\
3/16-1/16\,\sqrt {6}&1/4-1/16\,\sqrt {6}&1/8-1/16\,\sqrt {2}&1/8\\
3/16-1/16\,\sqrt {6}&1/4-1/16\,\sqrt {6}&1/8-1/16\,\sqrt {2}&1/8\\
5/16+1/16\,\sqrt {6}&1/4+1/16\,\sqrt {6}&3/8+1/16\,\sqrt {2}&3/8
\end {array} \right|\right|.
\end{displaymath}

These states are separable and they satisfy the Bell-CHSH inequality \cite{Qubit portrait of qudit states and Bell inequalities}.
\subsection{Qubit-qutrit system}
Some Bell-type numbers as functions of arbitrary correlation matrix $\hat{C}_{2\otimes3}$ are organized in the Table $5$. These numbers are the values of the corresponding Bell-CHSH form at given vertices constructed by employing permutations of probability vectors $\textbf{x},\textbf{y},\textbf{z},\textbf{t}$.
\begin{table}[h]
  \centering
\begin{tabular}{|c|c|}
  \hline
   Vertex  & Bell-type numbers\\
  \hline
  00000000 & $c_{16} + c_{26} + c_{56} + c_{46} + c_{36} + c_{66}$ \\
  00000001 & $c_{16} + c_{35} + c_{26} + c_{56} + c_{46} + c_{65}$ \\
  01000000 & $c_{16} + c_{26} + c_{53} + c_{43} + c_{36} + c_{63}$ \\
  01100000 & $c_{14} + c_{26} + c_{53} + c_{41} + c_{36} + c_{63}$ \\
  01010101 & $c_{15} + c_{35} + c_{25} + c_{52} + c_{42} + c_{62}$ \\
  10000000 & $c_{13} + c_{33} + c_{23} + c_{56} + c_{46} + c_{66}$ \\
  10101010 & $c_{21} + c_{11} + c_{31} + c_{54} + c_{44} + c_{64}$ \\
  11000000 & $c_{13} + c_{33} + c_{23} + c_{53} + c_{43} + c_{63}$ \\
  11100000 & $c_{11} + c_{33} + c_{23} + c_{53} + c_{41} + c_{63}$ \\
    \hline
\end{tabular}
\caption{The Bell-type numbers of qubit-qutrit system}
\end{table}

The Bell-CHSH form for qubit-qutrit system is the function:
\begin{gather*}
B_{2\otimes3}(\textbf{x},\textbf{y},\textbf{z},\textbf{t})=4+4\,x_{{1}}z_{{1}}-2\,x_{{2}}t_{{2}}+4\,x_{{1}}y_{{1}}+2\,x_{{1}}y_{{2}}+4\,x_{{2}}z_{{1}}+4\,x_{{2}}y_{{1}}+2\,x_{{2}}z_{{2}}+\\
+2\,x_{{1}}z_{{2}}+2\,x_{{1}}t_{{2}}+4\,x_{{1}}t_{{1}}-6\,x_{{1}}-2\,x_{{2}}-4\,y_{{1}}-4\,z_{{1}}-2\,y_{{2}}-2\,z_{{2}}+2\,x_{{2}}y_{{2}}-4\,x_{{2}}t_{{1}}
\end{gather*}

The quantum Bell-type numbers are calculated by exploiting two unitary matrices from $SU(2)$ for the qubit and three unitary matrices from $SU(3)$ for the qutrit. This produces six quantum probability distributions entering the final expression for the stochastic matrix associated with the entangled state $\hat{\rho}_{2\otimes3}$. This way the first Euler angles violating the classical local reality inequalities(Bell-type inequalities) were discovered. The Bell-CHSH function in quantum mechanics is:
\begin{gather*}
B_{2\otimes3}(\theta_1,\theta_2,\theta_3,\theta_4,\theta_5)=
\cos\!\left(\mathrm{{\theta}_1}\right)\, \cos\!\left(\mathrm{{\theta}_3}\right) + \cos\!\left(\mathrm{{\theta}_1}\right)\, \cos\!\left(\mathrm{{\theta}_4}\right) + \cos\!\left(\mathrm{{\theta}_2}\right)\, \cos\!\left(\mathrm{{\theta}_3}\right) + \\
\cos\!\left(\mathrm{{\theta}_1}\right)\, \cos\!\left(\mathrm{{\theta}_5}\right) + \cos\!\left(\mathrm{{\theta}_2}\right)\, \cos\!\left(\mathrm{{\theta}_4}\right) - \cos\!\left(\mathrm{{\theta}_2}\right)\, \cos\!\left(\mathrm{{\theta}_5}\right)
\end{gather*}

Table $6$ summarizes our current results for maximal violation of Bell-type inequalities in qubit-qutrit system. The upper bound for Bell-type inequality for the qubit-qutrit system was obtained by choosing two directions $1$ and $2$ for qubit and three directions $1'$,$2'$,$3'$ for the qutrit.

\begin{table}[h]
  \centering
\begin{tabular}{|c|c|c|c|c|c|}
  \hline
  \multirow{2}{*}{Angles}    & \multicolumn{5}{|c|}{Directions} \\
  \hline
            & 1 & 2 & 1' & 2' & 3' \\
  \hline
  $\varphi$ & $0.0004$ & $3.1450$  & $0$ & $0$ & $0$ \\
  \hline
  $\theta$  & $0.2368$ & $0.2538$  & $3.1471$ & $3.1292$ & $1.6024$ \\
  \hline
           & \multicolumn{5}{|c|}{$B_{2\otimes3}=4.0612$} \\
  \hline
\end{tabular}
\caption{A stronger violation of Bell-CHSH inequality for qubit-qutrit system}
\end{table}

\subsection{Two-qutrit system}
The stochastic matrix $\hat{M}_{\mathbf{x},\mathbf{y},\mathbf{z}, \mathbf{r},\mathbf{s},\mathbf{t}}$ associated with the composite system of two trits is taken in the form of $(10)$. The elements of special sign matrix $\hat{\sigma}_{3\otimes3}$ are received by arranging the signs of probabilities entering the average value
\begin{displaymath}
\langle m_1m_2\rangle_{\mathbf{n}_1,\mathbf{n}_2}=\sum_{m_1,m_2=\pm1,0} w\left(m_1,m_2,\mathbf{n}_1,\mathbf{n}_2\right)m_1m_2
\end{displaymath}
into columns of the corresponding sign matrix:
\begin{displaymath}
\hat{\sigma}_{3\otimes3}=\left|\left|
\begin {array}{rrrrrrrrr}
1&1&1&1&1&1&1&1&1\\
0&0&0&0&0&0&0&0&0\\
-1&-1&-1&-1&-1&-1&-1&-1&-1\\
0&0&0&0&0&0&0&0&0\\
0&0&0&0&0&0&0&0&0\\
0&0&0&0&0&0&0&0&0\\
-1&-1&-1&-1&-1&-1&-1&-1&-1\\
0&0&0&0&0&0&0&0&0\\
1&1&1&1&1&1&1&1&1
\end {array} \right|\right|
\end{displaymath}
The Bell-CHSH form in this case reads:
\begin{gather*}
B_{3\otimes3}(\textbf{x},\textbf{y},\textbf{z},\textbf{r},\textbf{s},\textbf{t})=
7-6\,x_{{1}}-6\,y_{{1}}-2\,z_{{1}}-3\,x_{{2}}-3\,y_{{2}}-z_{{2}}-6\,r_{{1}}-6\,s_{{1}}-2\,t_{{1}}-3\,r_{{2}}-3\,s_{{2}}-t_{{2}}+\\
+4\,z_{{1}}s_{{1}}-4\,z_{{1}}t_{{1}}+2\,z_{{1}}s_{{2}}-z_{{2}}t_{{2}}-2\,z_{{2}}t_{{1}}+z_{{2}}s_{{2}}+2\,z_{{2}}s_{{1}}+z_{{2}}r_{{2}}+2\,z_{{2}}r_{{1}}++4\,x_{{1}}r_{{1}}+2\,x_{{1}}r_{{2}}+\\
+4\,x_{{1}}s_{{1}}+2\,x_{{1}}s_{{2}}+4\,x_{{1}}t_{{1}}+2\,x_{{1}}t_{{2}}+2\,x_{{2}}r_{{1}}+x_{{2}}r_{{2}}+2\,x_{{2}}s_{{1}}+x_{{2}}s_{{2}}+2\,x_{{2}}t_{{1}}+x_{{2}}t_{{2}}+4\,y_{{1}}r_{{1}}+\\
+2\,y_{{1}}r_{{2}}+4\,y_{{1}}s_{{1}}+2\,y_{{1}}s_{{2}}+4\,y_{{1}}t_{{1}}+2\,y_{{1}}t_{{2}}+2\,y_{{2}}r_{{1}}+y_{{2}}r_{{2}}+2\,y_{{2}}s_{{1}}\\
+y_{{2}}s_{{2}}+2\,y_{{2}}t_{{1}}+y_{{2}}t_{{2}}+4\,z_{{1}}r_{{1}}+2\,z_{{1}}r_{{2}}-2\,z_{{1}}t_{{2}}
\end{gather*}

Some of Bell-type numbers for qutrit-qutrit system are gathered in Table $7$. These numbers were obtained by evaluating the Bell-CHSH form $B_{3\otimes3}(\textbf{x},\textbf{y},\textbf{z},\textbf{r},\textbf{s},\textbf{t})$ at all-possible vertices constituted by permutations of $\textbf{x},\textbf{y},\textbf{z},\textbf{r},\textbf{s},\textbf{t}$ probability vectors. Using the bilinearity of the Bell-CHSH form one can say with certainty that to obtain the upper bound of Bell-type inequality one needs to simply compare these numbers for a given matrix $\hat{C}_{3\otimes3}$.
\begin{table}[h]
  \centering
\begin{tabular}{|c|c|}
  \hline
   Vertex & Bell-type numbers \\
  \hline
  000000000000 & $c_{49}+c_{59}+c_{69}+c_{79}+c_{89}+c_{99}+c_{19}+c_{29}+c_{39}$ \\
  000000000001 & $c_{49}+c_{59}+c_{68}+c_{79}+c_{89}+c_{98}+c_{19}+c_{29}+c_{38}$ \\
  100000000000 & $c_{49}+c_{59}+c_{69}+c_{79}+c_{89}+c_{99}+c_{13}+c_{23}+c_{33}$ \\
  010101010101 & $c_{45}+c_{55}+c_{65}+c_{75}+c_{85}+c_{95}+c_{15}+c_{25}+c_{35}$ \\
  101010101010 & $c_{51}+c_{61}+c_{71}+c_{81}+c_{91}+c_{11}+c_{21}+c_{31}+c_{41}$ \\

    \hline
\end{tabular}
\caption{The Bell-type numbers of qutrit-qutrit system}
\end{table}

This way the several numeric values of Bell-type numbers corresponding to the particular case of matrix $\hat{C}_{3\otimes3}=\hat{I}_{3\otimes3}$ can be obtained:
\begin{displaymath}
B=-9,-3,0,6,9,15,18,6,0,-12,-18,-30,-36,0,-6,0,3,9,12,18,12,0,-6,-18,-24,-12,-6,-9
\end{displaymath}
The values $\pm12$, $\pm15$, $\pm18$, $\pm24$, $\pm30$, $\pm36$ are non-physical as they correspond to fictitious vertices (see page $9$) and should therefore be discarded. The corresponding quantum Bell-type numbers are received by using six matrices from $SU(3)$ three per each of the correlating qutrits. The resulting $9\times9$ stochastic matrices enter the expression for Bell-type numbers along with the special sign matrix $\hat{I}_{3\otimes3}$. These expressions were evaluated at various values of Euler angles in the $12$-dimensional phase space. The Bell-CHSH form corresponding to sign matrix $\hat{I}_{3\otimes3}$ has the form:
\begin{gather*}
B_{3\otimes3}(\theta_1,\theta_2,\theta_3,\theta_4,\theta_5,\theta_6)=\frac{2\, \cos\!\left(\mathrm{{\theta}_1} - \mathrm{{\theta}_4}\right)}{3} + \frac{2\, \cos\!\left(\mathrm{{\theta}_1} - \mathrm{{\theta}_5}\right)}{3} + \frac{2\, \cos\!\left(\mathrm{{\theta}_2} - \mathrm{{\theta}_4}\right)}{3} + \frac{2\, \cos\!\left(\mathrm{{\theta}_1} - \mathrm{{\theta}_6}\right)}{3} + \\
+ \frac{2\, \cos\!\left(\mathrm{{\theta}_2} - \mathrm{{\theta}_5}\right)}{3} + \frac{2\, \cos\!\left(\mathrm{{\theta}_3} - \mathrm{{\theta}_4}\right)}{3}+ \frac{2\, \cos\!\left(\mathrm{{\theta}_2} - \mathrm{{\theta}_6}\right)}{3} + \frac{2\, \cos\!\left(\mathrm{{\theta}_3} - \mathrm{{\theta}_5}\right)}{3} - \frac{2\, \cos\!\left(\mathrm{{\theta}_3} - \mathrm{{\theta}_6}\right)}{3}
\end{gather*}
It is notable that this B-form does not depend on angles $\varphi_i,i=1,2,3,4,5,6$ for the chosen form of the matrix $\hat{I}_{3\otimes3}$ and predicts a maximum value $B_{upper}=6<7$, i.e. the Bell-type inequality is preserved for the particular choice of matrix $\hat{C}_{3\otimes3}$.

\subsection{Three-qubit system}
We start with following stochastic matrices built-up from probability distributions corresponding to each of the $2$-level subsystems in the three-bit system:
\begin{displaymath}
\hat{M}_{\mathbf{x}}=
\left|\left|\begin {array}{cc} x_{{1}}&x_{{2}}\\\noalign{\medskip}1-x_{{1}}&1-x_{{2}}\end {array} \right|\right|,
\qquad \hat{M}_{\mathbf{y}}=\left|\left| \begin {array}{cc} y_{{1}}&y_{{2}}\\\noalign{\medskip}1-y_{{1}}&1-y_{{2}}\end {array} \right|\right|,
\qquad \hat{M}_{\mathbf{z}}=\left|\left| \begin {array}{cc} z_{{1}}&z_{{2}}\\\noalign{\medskip}1-z_{{1}}&1-z_{{2}}\end {array} \right|\right|
\end{displaymath}

The resulting $8\times8$ stochastic matrix of the composite system $\hat{M}_{\mathbf{x},\mathbf{y},\mathbf{z}}=\hat{M}_\mathbf{x}\otimes\hat{M}_\mathbf{y}\otimes\hat{M}_\mathbf{z}$  has the following elements entering its columns $x_1y_1z_1,x_1y_1(1-z_1)$, $x_1(1-y_1) z_1$, $x_1(1-y_1)(1-z_1)$, $(1-x_1)y_1z_1$, $(1-x_1)y_1(1-z_1)$, $(1-x_1)(1-y_1) z_{{1}}$, $(1-x_1)(1-y_1)(1-z_{{1}})$. The second and other columns have the same structure obtained with cyclic permutation of $\mathbf{x}$, $\mathbf{y}$, $\mathbf{z}$.

The Bell-CHSH form in the framework of classical theory of probability reads:
\begin{gather*}
B_{2\otimes2\otimes2}(\textbf{x},\textbf{y},\textbf{z})=
-6+8x_{2}y_{1}z_{1}+8x_{1}y_{2}z_{2}+8x_{2}y_{2}z_{1}-8x_{2}y_{2}z_{2}-8x_{1}z_{1}+8x_{2}y_{1}z_{2}-\\
-8y_{1}z_{2}-8y_{1}z_{1}+8x_{1}y_{2}z_{1}-8x_{2}z_{1}-8x_{1}z_{2}-8x_{2}y_{1}-8x_{1}y_{2}+ \\
+8x_{1}+4x_{2}+8y_{1}+4y_{2}+8z_{1}+4z_{2}-8y_{1}x_{1}+8x_{1}y_{1}z_{1}+8x_{1}y_{1}z_{2}-8y_{2}z_{1}
\end{gather*}
The calculation for the first $14$ vertices via arbitrary matrix $\hat{C}_{2\otimes2\otimes2}$ yields to following Bell-type numbers gathered in Table $8$.
\begin{table}[h]
  \centering
\begin{tabular}{|ccc|c|}
  \hline
     & Vertex  &    & Bell-type numbers \\
  \hline
  \textbf{x}  & \textbf{y}  & \textbf{z}  &  \\
  \hline
  00 & 00 & 00 & $c_{18} + c_{28} + c_{38} + c_{48} + c_{58} + c_{68} + c_{78} + c_{88}$ \\
  10 & 00 & 00 & $c_{14} + c_{24} + c_{34} + c_{44} + c_{58} + c_{68} + c_{78} + c_{88}$ \\
  11 & 00 & 00 & $c_{14} + c_{24} + c_{34} + c_{44} + c_{54} + c_{64} + c_{74} + c_{84}$ \\
  11 & 10 & 00 & $c_{12} + c_{22} + c_{34} + c_{44} + c_{52} + c_{62} + c_{74} + c_{84}$ \\
  11 & 11 & 00 & $c_{12} + c_{22} + c_{32} + c_{42} + c_{52} + c_{62} + c_{72} + c_{82}$ \\
  11 & 11 & 10 & $c_{11} + c_{22} + c_{31} + c_{42} + c_{51} + c_{62} + c_{71} + c_{82}$ \\
  11 & 11 & 11 & $c_{11} + c_{21} + c_{31} + c_{41} + c_{51} + c_{61} + c_{71} + c_{81}$ \\
  01 & 01 & 01 & $c_{18} + c_{27} + c_{36} + c_{45} + c_{54} + c_{63} + c_{72} + c_{81}$ \\
  10 & 10 & 10 & $c_{11} + c_{22} + c_{33} + c_{44} + c_{55} + c_{66} + c_{77} + c_{88}$ \\
  00 & 00 & 01 & $c_{18} + c_{27} + c_{38} + c_{47} + c_{58} + c_{67} + c_{78} + c_{87}$ \\
  00 & 00 & 11 & $c_{17} + c_{27} + c_{37} + c_{47} + c_{57} + c_{67} + c_{77} + c_{87}$ \\
  00 & 01 & 11 & $c_{17} + c_{27} + c_{35} + c_{45} + c_{57} + c_{67} + c_{75} + c_{85}$ \\
  00 & 11 & 11 & $c_{15} + c_{25} + c_{35} + c_{45} + c_{55} + c_{65} + c_{75} + c_{85}$ \\
  01 & 11 & 11 & $c_{15} + c_{25} + c_{35} + c_{45} + c_{51} + c_{61} + c_{71} + c_{81}$ \\
  \hline
\end{tabular}
\caption{The Bell-type numbers of three-qubit system}
\end{table}

The selection of correlation matrix
\begin{displaymath}
\hat{C}_{2\otimes2\otimes2}=\hat{I}_{2\otimes2\otimes2}=\left|\left| \begin {array}{cccccccc} 1&-1&-1&1&-1&1&1&-1
\\\noalign{\medskip}1&-1&-1&1&-1&1&1&-1\\\noalign{\medskip}1&-1&-1&1&-
1&1&1&-1\\\noalign{\medskip}1&-1&-1&1&-1&1&1&-1\\\noalign{\medskip}1&-
1&-1&1&-1&1&1&-1\\\noalign{\medskip}1&-1&-1&1&-1&1&1&-1
\\\noalign{\medskip}1&-1&-1&1&-1&1&1&-1\\\noalign{\medskip}-1&1&1&-1&1
&-1&-1&1\end {array}
 \right|\right|
\end{displaymath}
as well as choices with sign inversions determine the upper bound of the system by comparing all Bell-type numbers. The upper bound is $6$.

The quantum correlations are taken into consideration via the simplest entangled state

\begin{displaymath}
\hat{\rho}_{2\otimes2\otimes2}=\dfrac 12\left|\left|%
\begin{array}{ccccccccc}
1 & 0 & 0 & 0 & 0 & 0 & 0 & 1\\
0 & 0 & 0 & 0 & 0 & 0 & 0 & 0\\
0 & 0 & 0 & 0 & 0 & 0 & 0 & 0\\
0 & 0 & 0 & 0 & 0 & 0 & 0 & 0\\
0 & 0 & 0 & 0 & 0 & 0 & 0 & 0\\
0 & 0 & 0 & 0 & 0 & 0 & 0 & 0\\
0 & 0 & 0 & 0 & 0 & 0 & 0 & 0\\
1 & 0 & 0 & 0 & 0 & 0 & 0 & 1\\
 \end{array}%
\right|\right|
\end{displaymath}
The Bell-CHSH form for the three-qubit system reads:
\begin{gather*}
B_{2\otimes2\otimes2}(\theta_1,\theta_3,\theta_3,\theta_4,\theta_5,\theta_6,{\varphi}_1,{\varphi}_2,{\varphi}_3,{\varphi}_4,{\varphi}_5, {\varphi}_6) = \\
=8\, \cos\!\left( - \mathrm{{\varphi}_2} - \mathrm{{\varphi}_4} - \mathrm{{\varphi}_6}\right)\, \cos\!\left(\frac{\mathrm{{\theta}_2}}{2}\right)\, \cos\!\left(\frac{\mathrm{{\theta}_4}}{2}\right)\, \cos\!\left(\frac{\mathrm{{\theta}_6}}{2}\right)\, \sin\!\left(\frac{\mathrm{{\theta}_2}}{2}\right)\, \sin\!\left(\frac{\mathrm{{\theta}_4}}{2}\right)\, \sin\!\left(\frac{\mathrm{{\theta}_6}}{2}\right) - \\
-8\, \cos\!\left( - \mathrm{{\varphi}_1} - \mathrm{{\varphi}_3} - \mathrm{{\varphi}_6}\right)\, \cos\!\left(\frac{\mathrm{{\theta}_1}}{2}\right)\, \cos\!\left(\frac{\mathrm{{\theta}_3}}{2}\right)\, \cos\!\left(\frac{\mathrm{{\theta}_6}}{2}\right)\, \sin\!\left(\frac{\mathrm{{\theta}_1}}{2}\right)\, \sin\!\left(\frac{\mathrm{{\theta}_3}}{2}\right)\, \sin\!\left(\frac{\mathrm{{\theta}_6}}{2}\right) - \\
-8\, \cos\!\left( - \mathrm{{\varphi}_1} - \mathrm{{\varphi}_4} - \mathrm{{\varphi}_5}\right)\, \cos\!\left(\frac{\mathrm{{\theta}_1}}{2}\right)\, \cos\!\left(\frac{\mathrm{{\theta}_4}}{2}\right)\, \cos\!\left(\frac{\mathrm{{\theta}_5}}{2}\right)\, \sin\!\left(\frac{\mathrm{{\theta}_1}}{2}\right)\, \sin\!\left(\frac{\mathrm{{\theta}_4}}{2}\right)\, \sin\!\left(\frac{\mathrm{{\theta}_5}}{2}\right) - \\
-8\, \cos\!\left( - \mathrm{{\varphi}_2} - \mathrm{{\varphi}_3} - \mathrm{{\varphi}_5}\right)\, \cos\!\left(\frac{\mathrm{{\theta}_2}}{2}\right)\, \cos\!\left(\frac{\mathrm{{\theta}_3}}{2}\right)\, \cos\!\left(\frac{\mathrm{{\theta}_5}}{2}\right)\, \sin\!\left(\frac{\mathrm{{\theta}_2}}{2}\right)\, \sin\!\left(\frac{\mathrm{{\theta}_3}}{2}\right)\, \sin\!\left(\frac{\mathrm{{\theta}_5}}{2}\right) - \\
-8\, \cos\!\left( - \mathrm{{\varphi}_1} - \mathrm{{\varphi}_4} - \mathrm{{\varphi}_6}\right)\, \cos\!\left(\frac{\mathrm{{\theta}_1}}{2}\right)\, \cos\!\left(\frac{\mathrm{{\theta}_4}}{2}\right)\, \cos\!\left(\frac{\mathrm{{\theta}_6}}{2}\right)\, \sin\!\left(\frac{\mathrm{{\theta}_1}}{2}\right)\, \sin\!\left(\frac{\mathrm{{\theta}_4}}{2}\right)\, \sin\!\left(\frac{\mathrm{{\theta}_6}}{2}\right) - \\
-8\, \cos\!\left( - \mathrm{{\varphi}_2} - \mathrm{{\varphi}_3} - \mathrm{{\varphi}_6}\right)\, \cos\!\left(\frac{\mathrm{{\theta}_2}}{2}\right)\, \cos\!\left(\frac{\mathrm{{\theta}_3}}{2}\right)\, \cos\!\left(\frac{\mathrm{{\theta}_6}}{2}\right)\, \sin\!\left(\frac{\mathrm{{\theta}_2}}{2}\right)\, \sin\!\left(\frac{\mathrm{{\theta}_3}}{2}\right)\, \sin\!\left(\frac{\mathrm{{\theta}_6}}{2}\right) - \\
-8\, \cos\!\left( - \mathrm{{\varphi}_2} - \mathrm{{\varphi}_4} - \mathrm{{\varphi}_5}\right)\, \cos\!\left(\frac{\mathrm{{\theta}_2}}{2}\right)\, \cos\!\left(\frac{\mathrm{{\theta}_4}}{2}\right)\, \cos\!\left(\frac{\mathrm{{\theta}_5}}{2}\right)\, \sin\!\left(\frac{\mathrm{{\theta}_2}}{2}\right)\, \sin\!\left(\frac{\mathrm{{\theta}_4}}{2}\right)\, \sin\!\left(\frac{\mathrm{{\theta}_5}}{2}\right) - \\
-8\, \cos\!\left( - \mathrm{{\varphi}_1} - \mathrm{{\varphi}_3} - \mathrm{{\varphi}_5}\right)\, \cos\!\left(\frac{\mathrm{{\theta}_1}}{2}\right)\, \cos\!\left(\frac{\mathrm{{\theta}_3}}{2}\right)\, \cos\!\left(\frac{\mathrm{{\theta}_5}}{2}\right)\, \sin\!\left(\frac{\mathrm{{\theta}_1}}{2}\right)\, \sin\!\left(\frac{\mathrm{{\theta}_3}}{2}\right)\, \sin\!\left(\frac{\mathrm{{\theta}_5}}{2}\right).
\end{gather*}

We hope to find the deviation from the classical bound for Bell-type number due to quantum correlations between three-qubits in our next article. The finding of these deviation is complicated by the necessity of choosing several scenarios of pair interactions inside the system. These interactions in form of remarkable correlations between dynamic variables are absorbed in a single matrix $\hat{C}$ with several physical constraints \cite{Bell-type inequalities in classical probability theory}. The three-qubit B-form obtained above in principle permits violation of Bell-type inequality. In that perspective in our approach the sign matrix $\hat{I}_{2\otimes2\otimes2}$ is much ``better chosen'' than $\hat{I}_{3\otimes3}$ for the two-qutrit system.

\section*{Acknowledgments}
The study was supported by the Russian Foundation for Basic Research under Projects Nos. $07-02-00598$ and $09-02-00142$.

\addcontentsline{toc}{section}{
\end{document}